\documentstyle[12pt]{article}
\begin{document}
\begin{flushright}
\baselineskip=12pt
DOE/ER/40717--32\\
CTP-TAMU-32/96\\
ACT-11/96\\
\tt hep-ph/9608275
\end{flushright}

\begin{center}
\vglue 1.5cm
{\Large\bf Experimental consequences of no-scale supergravity in light of the CDF $ee\gamma\gamma$ event}
\vglue 1.5cm
{\Large Jorge L. Lopez$^1$ and D.V. Nanopoulos$^{2,3}$}
\vglue 1cm
\begin{flushleft}
$^1$Department of Physics, Bonner Nuclear Lab, Rice University\\ 6100 Main
Street, Houston, TX 77005, USA\\
$^2$Center for Theoretical Physics, Department of Physics, Texas A\&M
University\\ College Station, TX 77843--4242, USA\\
$^3$Astroparticle Physics Group, Houston Advanced Research Center (HARC)\\
The Mitchell Campus, The Woodlands, TX 77381, USA\\
\end{flushleft}
\end{center}

\vglue 1cm
\begin{abstract}
We explore two possible interpretations for the CDF $ee\gamma\gamma+E_T\hskip-13pt/\quad$ event in the context of a recently proposed one-parameter no-scale supergravity model with a light gravitino. We delineate the region in parameter space consistent with the kinematics of the event interpreted either as selectron pair-production 
($p\bar p\to\tilde e^+\tilde e^-\,X$, $\tilde e=\tilde e_R,\tilde e_L$) or as chargino pair-production ($p\bar p\to\chi^+_1\chi^-_1\,X$). In the context of this model, the selectron interpretation requires $\tilde e=\tilde e_R$ and predicts comparable rates for $\ell^\pm\gamma\gamma+E_T\hskip-13pt/\quad$, whereas the chargino interpretation predicts comparable rates for $(\ell^\pm\ell^{'+}\ell^{'-},\ell^+\ell^-,\ell^+\ell^-jj)\,\gamma\gamma
+E_T\hskip-13pt/\quad$. We also consider the constraints from LEP~1.5 and the
expectations for LEP~2. We point out that one of the acoplanar photon pairs observed by the OPAL collaboration at LEP~1.5 may be attributable to supersymmetry in the present model via $e^+e^-\to\chi^0_1\chi^0_1\to
\gamma\gamma+E_T\hskip-13pt/\quad$. We also show how similar future events at
LEP~2 may be used to deduce the neutralino mass.
\end{abstract}
\vspace{1cm}
\begin{flushleft}
\baselineskip=12pt
August 1996\\
\medskip
{\small\tt lopez@physics.rice.edu}\\
{\small\tt dimitri@phys.tamu.edu}
\end{flushleft}
\newpage
\setcounter{page}{1}
\pagestyle{plain}
\baselineskip=14pt

\section{Introduction}
\label{sec:introduction}
After the stunning confirmation of the Standard Model predictions at the 
Tevatron and LEP~1, the field of contenders for the much anticipated `new
physics' has, in the minds of most, been narrowed down to just one: supersymmetry.  However, the many years elapsed since the invention of supersymmetry have produced abundant theoretical speculation as to the specific nature of supersymmetry, its origins, its breaking, and the spectrum of superparticles. Recent observations at the Tevatron, in the form of a puzzling $ee\gamma\gamma+E_T\hskip-13pt/\quad$ event \cite{Park}, appear to indicate that experiment may have finally reached the sensitivity required to observe the first direct manifestations of supersymmetry \cite{DinePRL,KanePRL,Gravitino}. If this event is indeed the result of an underlying supersymmetric production process, as might be deduced from the observation of additional related events at the Tevatron or LEP~2, it will herald a new era in elementary particle physics. The {\em prima facie} evidence for supersymmetry contains the standard
missing-energy characteristic of supersymmetric production processes, but it
also contains a surprising hard-photon component (as far as the Minimal Supersymmetry Standard Model is concerned), which eliminates all conceivable
Standard Model backgrounds and may prove extremely discriminating among different models of low-energy supersymmetry.

The present supersymmetric explanations of the CDF event fall into two 
phenomenological classes: either the lightest neutralino ($\chi^0_1$) is the lightest supersymmetric particle, and the second-to-lightest neutralino decays radiatively to it at the one-loop level ($\chi^0_2\to\chi^0_1\gamma$); or the gravitino ($\widetilde G$) is the lightest supersymmetric particle, and the lightest neutralino decays radiatively to it at the tree level ($\chi^0_1\to\widetilde G\gamma$). The former `neutralino-LSP' scenario requires a configuration of gaugino masses that precludes the usual gaugino mass unification relation of unified models, although it can occur in some restricted region of the MSSM parameter space. The latter `gravitino-LSP' scenario requires only that the lightest neutralino has a photino component, as is typically the case in many supersymmetric models. The underlying process that leads to such final states has been suggested to be that of selectron pair-production ($q\bar q\to \widetilde e^+
\widetilde e^-$, $\widetilde e=\widetilde e_R,\widetilde e_L$), with subsequent decay $\widetilde e\to e\chi^0_2$ or $\widetilde e\to e\chi^0_1$ in the neutralino-LSP and gravitino-LSP scenarios respectively. In the gravitino-LSP scenario, the alternative possibility of chargino pair-production ($q\bar q\to\chi^+_1\chi^-_1$,$\chi^\pm_1\to e^\pm\nu_e\chi^0_1$) has also been briefly mentioned \cite{Kane} (this possibility is not available in the neutralino-LSP scenario \cite{KaneN}). 

Of these two classes of explanations, only the gravitino-LSP one has generated
model-building efforts that try to embed such a scenario into a more fundamental theory at higher mass scales. These more predictive theories
include low-energy gauge-mediated dynamical supersymmetry breaking \cite{DinePRL,Dine} and no-scale supergravity \cite{Gravitino}. In this paper
we study in detail the kinematics of the observed event in both the selectron
and chargino interpretations in the context of our recently proposed {\em one-parameter} no-scale supergravity model \cite{Gravitino}. We delineate the
regions in parameter space that are consistent with the kinematical information, and then consider the rates for the various underlying processes that may occur within such regions of parameter space. Non-observation of related signals ({\em e.g.}, $e\gamma\gamma+E_T\hskip-13pt/\quad$) that are predicted to occur at significantly higher rates than the presumably observed one ({\em i.e.}, $ee\gamma\gamma+E_T\hskip-13pt/\quad$) imposes important constraints on the parameter space of the model. It should be noted that the possibility of observing supersymmetry at the Tevatron via weakly-interacting production processes had been shown early on to be particularly favorable in the context of no-scale supergravity \cite{LNWZ}. We also consider the constraints from LEP~1.5 and the prospects for supersymmetric particle
detection at LEP161 and LEP190. We show that one of the acoplanar photon pairs observed by the OPAL collaboration at LEP~1.5 may be attributable to supersymmetry in the present model via $e^+e^-\to\chi^0_1\chi^0_1\to
\gamma\gamma+E_T\hskip-13pt/\quad$. We should remark that because of the rather
restrictive nature of our one-parameter model, our experimental predictions
are unambiguous and highly correlated.

This paper is organized as follows. In Section~\ref{sec:model} we review
the basic model predictions and expand on the motivation for our model and its significance in the context of no-scale supergravity, flipped SU(5), and extended supergravities. In Sec.~\ref{sec:events} we consider the CDF event
from the perspective of the selectron and chargino interpretation, delineate
the allowed regions in parameter space, and contrast the rates for related
processes that might have been observed. In Sec.~\ref{sec:LEP} we determine
the constraints from LEP~1.5, study in some detail the OPAL acoplanar photon
events, and explore the prospects for particle detection at LEP~2. In Sec.~\ref{sec:conclusions} we summarize our conclusions.

\section{The model}
\label{sec:model}
\subsection{Motivation}
\label{sec:motivation}
Supergravity models are described in terms of two functions, the K\"ahler
function $G=K+\ln|W|^2$, where $K$ is the K\"ahler potential and $W$ the
superpotential; and the gauge kinetic function $f$. Specification of these
functions determines the supergravity interactions and the soft supersymmetry-breaking parameters that arise after spontaneous breaking of
supergravity, which is parametrized by the gravitino mass $m_{3/2}=e^{K/2}\,|W|$. In standard supergravity scenarios the choices for $G$ and $f$ may follow from symmetry considerations or may be calculated in certain weakly-coupled heterotic string vacua (such as orbifolds or free-fermionic constructions). In these cases one generically obtains soft-supersymmetry-breaking parameters ({\em i.e.}, scalar and gaugino masses and scalar interactions) that are comparable to the gravitino mass: $m_0,m_{1/2},A_0\sim m_{3/2}$, although the specific proportionality coefficients in these relations may vary in magnitude and may even be non-universal. In the context of unified models, such standard scenarios do not appear to be consistent with a supersymmetric explanation of the CDF $ee\gamma\gamma+E_T\hskip-13pt/\quad$ event because, either the implied gaugino mass unification is violated (neutralino-LSP scenario), or the required light gravitino mass ({\em i.e.}, $m_{3/2}\ll1\,{\rm GeV}$) would render all supersymmetric particles comparably light (gravitino-LSP scenario). 

In the gravitino-LSP scenario, the main issue is that of {\em decoupling} the breaking of local supersymmetry (parametrized by $m_{3/2}$) from the breaking of global supersymmetry (parametrized by $m_0,m_{1/2}$). This decoupling is achieved {\em naturally} in the context of no-scale supergravity
\cite{no-scale,Lahanas}, where a judicious choice of the K\"ahler potential
($K$) yields $m_0=0m_{3/2}$ (and also $A_0=B_0=0$), which allows a very wide range of $m_{3/2}$ values. (No-scale supergravity is also obtained in 
weakly-coupled string models, as was shown early on by Witten \cite{Witten} and has more recently been explored in great detail in Ref.~\cite{LN}.) Without this (scalar-sector) decoupling, sizeable values of $m_0$ ({\em i.e.}, $m_0\sim M_Z$) cannot be obtained in the light gravitino scenario. As we discuss below, the selectron interpretation of the CDF event requires $m_0\ll M_Z$ \cite{Gravitino}, but there is no such requirement for the chargino interpretation. In this connection we should add that in unified models the possible values of $m_0/m_{1/2}$ are constrained by the proton decay rate via dimension-five operators. In the minimal SU(5) supergravity model one can show that $m_0/m_{1/2}>3$ is required \cite{pdecay}, thus rendering this decoupling
impossible. In contrast, in flipped SU(5) such operators are naturally suppressed, and $m_0=0$ is perfectly allowed \cite{revitalized}.

\subsection{A light gravitino}
The remaining and crucial question is the decoupling in the gaugino sector,
that depends on the choice of $f$, at least in traditional supergravity models.
The gaugino masses are given by
\begin{equation}
m_{1/2}=m_{3/2}\left({\partial_z f\over 2{\rm Re} f}\right)
\left({\partial_z G \over \partial_{z z^*} G}\right)\ ,
\label{eq:formula}
\end{equation}
where $z$ represents the hidden sector (moduli) fields in the model, and the
gaugino mass universality at the Planck scale is insured by a gauge-group
independent choice for $f$. As remarked above, the usual expressions for $f$ 
({\em e.g.}, in weakly-coupled string models) give $m_{1/2}\sim m_{3/2}$. This result is however avoided by considering the non-minimal choice $f\sim e^{-A z^q}$, where $A,q$ are constants \cite{EEN}. Assuming the standard no-scale expression $G=-3\ln(z+z^*)$, one can then readily show that \cite{EEN}
\begin{equation}
m_{1/2}\sim \left({m_{3/2}\over M}\right)^{1-{2\over3}q} M\ ,
\label{eq:result}
\end{equation}
where $M\approx10^{18}\,{\rm GeV}$ is the rescaled Planck mass.
The phenomenological requirement of $m_{1/2}\sim10^2\,{\rm GeV}$ then implies
${3\over4}\raisebox{-4pt}{$\,\stackrel{\textstyle{>}}{\sim}\,$} q
\raisebox{-4pt}{$\,\stackrel{\textstyle{>}}{\sim}\,$}
{1\over2}$ for $10^{-5}\,{\rm eV}
\raisebox{-4pt}{$\,\stackrel{\textstyle{<}}{\sim}\,$}
m_{3/2}\raisebox{-4pt}{$\,\stackrel{\textstyle{<}}{\sim}\,$}10^3\,{\rm eV}$.
Note that $q={3\over4}$ gives the relation $m_{3/2}\sim
m^2_{1/2}/M\sim10^{-5}\,{\rm eV}$, which was obtained very early on in
Ref.~\cite{BFN} from the perspective of hierarchical supersymmetry breaking in
extended N=8 supergravity. The recent theoretical impetus for supersymmetric
M-theory in 11 dimensions may also lend support to this result, as N=1 in D=11
corresponds to N=8 in D=4.

More specifically, the needed decoupling between the local and global breaking
of supersymmetry, that we have advocated above in order to insure a light
gravitino, appears to be realized in strongly coupled heterotic strings \cite{Horava}, which are best understood in terms of 11-dimensional supergravity. Moreover, the no-scale supergravity structure appears to also
emerge in strongly-coupled strings \cite{BD,Horava} and to take on a crucial role in suppressing the cosmological constant at the classical \cite{no-scale,Lahanas} and quantum levels \cite{Horava,LNprep}, as well as
avoiding flavor-changing neutral currents \cite{BD,LNprep}. These general remarks should sufficiently motivate our present phenomenological study.

\subsection{The spectrum}
Our model effectively fits within the usual no-scale supergravity models with universal soft-supersymmetry-breaking parameters given by
\begin{equation}
m_0=A_0=B_0=0\ .
\label{eq:parameters}
\end{equation}
When the constraints from radiative electroweak symmetry breaking are imposed,
the values of $|\mu|$ and $B(M_Z)$ are determined (for the fixed value of
$m_t=175\,{\rm GeV}$). Evolving the latter up to the Planck scale, and
demanding that it reproduce the boundary condition $B_0=0$, determines the
value of $\tan\beta$ in terms of the single parameter of the model ($m_{1/2}$).
One can show that the sign of $\mu$ is also determined by this procedure
\cite{One}. 

The precise low-energy spectrum of the model further depends on
the details of the evolution between the Planck scale and the electroweak
scale. One may envision a traditional scenario where unification occurs at the usual GUT scale ($M_{\rm LEP}\sim10^{16}\,{\rm GeV}$), while supersymmetry
breaking is communicated to the observable sector at the Planck scale. Alternatively one may supplement the MSSM spectrum with intermediate-scale
particles that delay unification up to the Planck scale \cite{sism}. A third
and perhaps more realistic scenario combines both approaches and has been recently realized in flipped SU(5) \cite{TwoStep}, achieving a partial unification at $M_{\rm LEP}$ [SU(2)$\times$SU(3)$\to$SU(5)] and string unification at $M_{Pl}$. However, this scenario depends on various parameters which make the analysis more complicated that needs to be at this stage. Here we opt for the second scenario, where unification occurs in a single step at the Planck scale. We expect that our numerical results will
remain qualitatively unchanged in any of these three possible unification scenarios.

\begin{figure}[p]
\vspace{6in}
\includegraphics{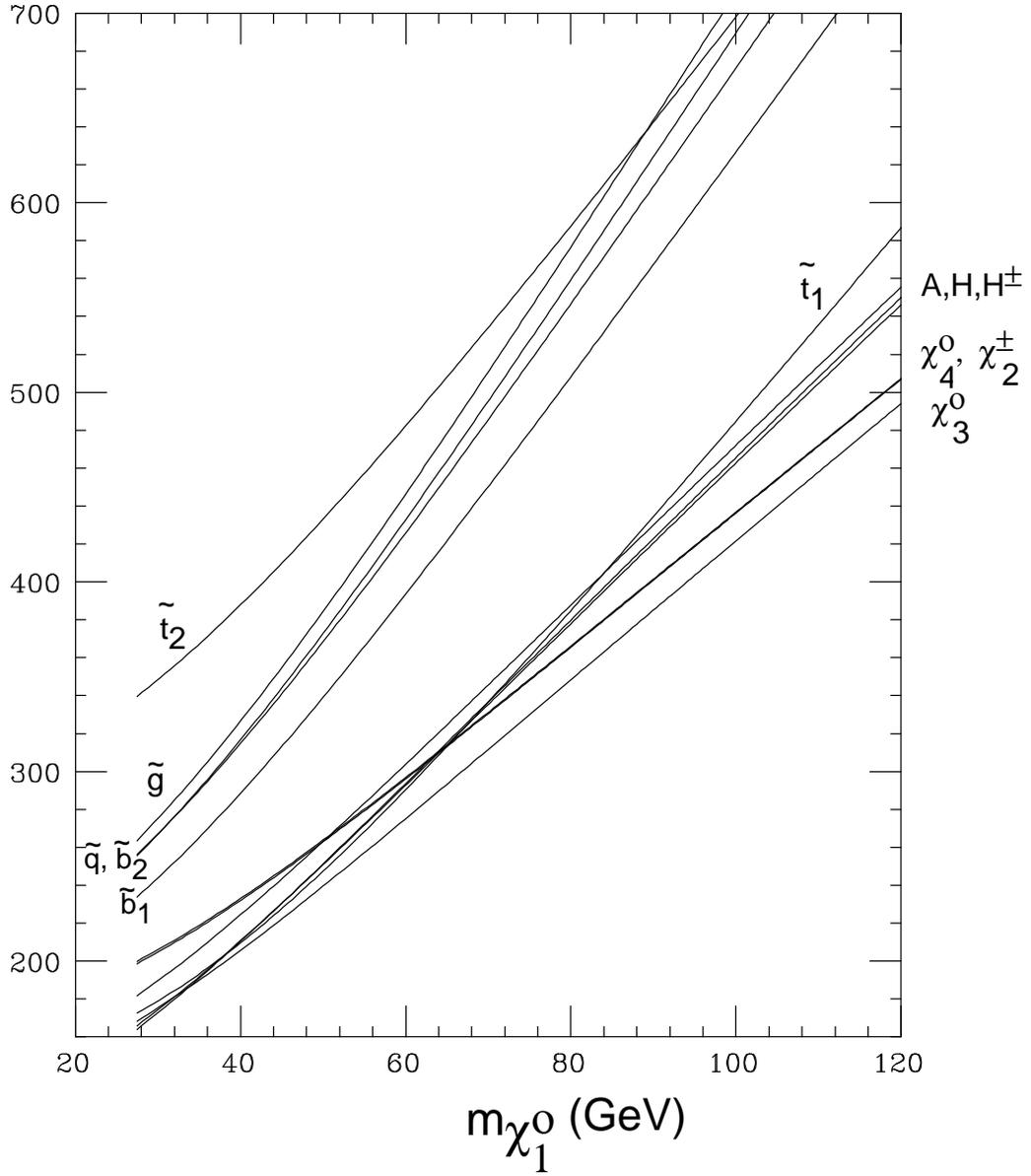}
\caption{The heavier members of the spectrum of our one-parameter model
versus the lightest neutralino mass. All masses in GeV. When multiple labels $a,b,c$ are attached to a particular line that may be split, the mass ordering
is $m_a\ge m_b\ge m_c$.}
\label{fig:heavy}
\end{figure}

The spectrum as a function of the lightest neutralino mass is given in
Fig.~\ref{fig:light} for the lighter particles (sleptons, lightest higgs boson,
lighter neutralinos and charginos) and in Fig.~\ref{fig:heavy} for the
heavier particles (gluino, squarks, heavy higgs bosons, heavier neutralinos and
charginos). The calculated value of $\tan\beta$ as a function of $m_{\chi^0_1}$ is displayed in Fig.~\ref{fig:light}. We also find that $|\mu|\approx m_{\chi^0_3}$, which can then be obtained from  Fig.~\ref{fig:heavy}. These figures show that the lightest neutralino (which is mostly bino) is always the next-to-lightest supersymmetric particle (NSLP), followed by the right-handed sleptons ($\tilde e_R,\tilde\mu_R,\tilde\tau_1$), the lighter neutralino/chargino ($\chi^0_2,\chi^\pm_1$), the sneutrino ($\tilde\nu$), and the left-handed sleptons ($\tilde e_L,\tilde\mu_L,\tilde\tau_2$). (The order of the second and third elements on this list is reversed for very light neutralino masses.) Note the splitting between the selectron/smuon masses and the stau mass due to the non-negligible value of the $\lambda_\tau$ Yukawa coupling. The lightest higgs boson crosses all sparticle lines and is in the range $m_h=(100-120)\,{\rm GeV}$. Also notable is that the average squark mass  ($\tilde q$) is slightly below the gluino mass ($\tilde g$) and the lightest top-squark ($\tilde t_1$) is somewhat lighter than both of these. 

As emphasized early on \cite{EEN}, the dominant decay of the lightest neutralino is via $\chi^0_1\to\gamma\widetilde G$, which has partial a width of \cite{EEN,Kane}
\begin{equation}
\Gamma(\chi^0_1\to\gamma\widetilde G)={\kappa_{1\gamma}\over48\pi}\,
{m^5_{\chi^0_1}\over (M_{Pl}\,m_{\widetilde G})^2}
=1.12\times10^{-11}\,{\rm GeV} \kappa_{1\gamma}
\left({m_{\chi^0_1}\over100\,{\rm GeV}}\right)^5
\left({m_{\widetilde G}\over 1\,{\rm eV}}\right)^{-2}\ ,
\label{eq:width}
\end{equation}
where $\kappa_{1\gamma}=|N_{11}\cos\theta_W+N_{12}\sin\theta_W|^2$ is the
square of the photino admixture of the lightest neutralino. In our model
$\kappa_{1\gamma}>0.5,0.6,0.7$ for $m_{\chi^0_1}>40,55,85\,{\rm GeV}$,
with a maximum value of 0.74, attained asymptotically when the lightest neutralino approaches a pure bino. The decay $\chi^0_1\to Z\widetilde G$,
accessible for $m_{\chi^0_1}>M_Z$, is greatly suppressed by the $\beta^8$
threshold behavior \cite{Kane}; we find 
$\Gamma(\chi^0_1\to Z\widetilde G)/\Gamma(\chi^0_1\to\gamma\widetilde G)<0.03$.
The decay $\chi^0_1\to h\widetilde G$ may also be accessible (for $m_{\chi^0_1}>120\,{\rm GeV}$) but it is completely negligible because of the
above kinematical suppression and because of the gaugino nature of the 
neutralino. In sum, we expect $B(\chi^0_1\to\gamma\widetilde G)\approx100\%$
throughout the whole parameter space. One can also consider the probability
for $\chi^0_1$ (with energy $E$) to travel a distance $x$ before decaying
$P(x)=1-e^{-x/L}$, where the decay length is given by \cite{Kane}
\begin{equation}
L=1.76\times10^{-3}\,(\kappa_{1\gamma})^{-1}
 \left(E^2/m^2_{\chi^0_1}-1\right)^{1/2}
\left({m_{\chi^0_1}\over100\,{\rm GeV}}\right)^{-5}
\left({m_{\widetilde G}\over 1\,{\rm eV}}\right)^2\,{\rm cm}\ .
\label{eq:length}
\end{equation}
The requirement $m_{3/2}\raisebox{-4pt}{$\,\stackrel{\textstyle{<}}{\sim}\,$}250\,{\rm eV}$ 
ensures a visible neutralino decay within the CDF (or any other such) detector \cite{KanePRL}.

\section{The CDF event}
\label{sec:events}
We now turn to the implications of the observed CDF $ee\gamma\gamma+E_T\hskip-13pt/\quad$ event, assuming that it is the result
of an underlying supersymmetric production process in the context of our
present model. The particulars of the event are listed in Table~\ref{Table1}.
We consider two interpretations (selectron and chargino
pair production), in each case delineating the region in parameter space
consistent with the observed kinematics of the event. We then determine the
rate for such processes and for related processes with similar signatures
that might have also been detected.

\subsection{Selectron interpretation}
\label{sec:selectron}
\begin{figure}[p]
\vspace{6in}
\includegraphics{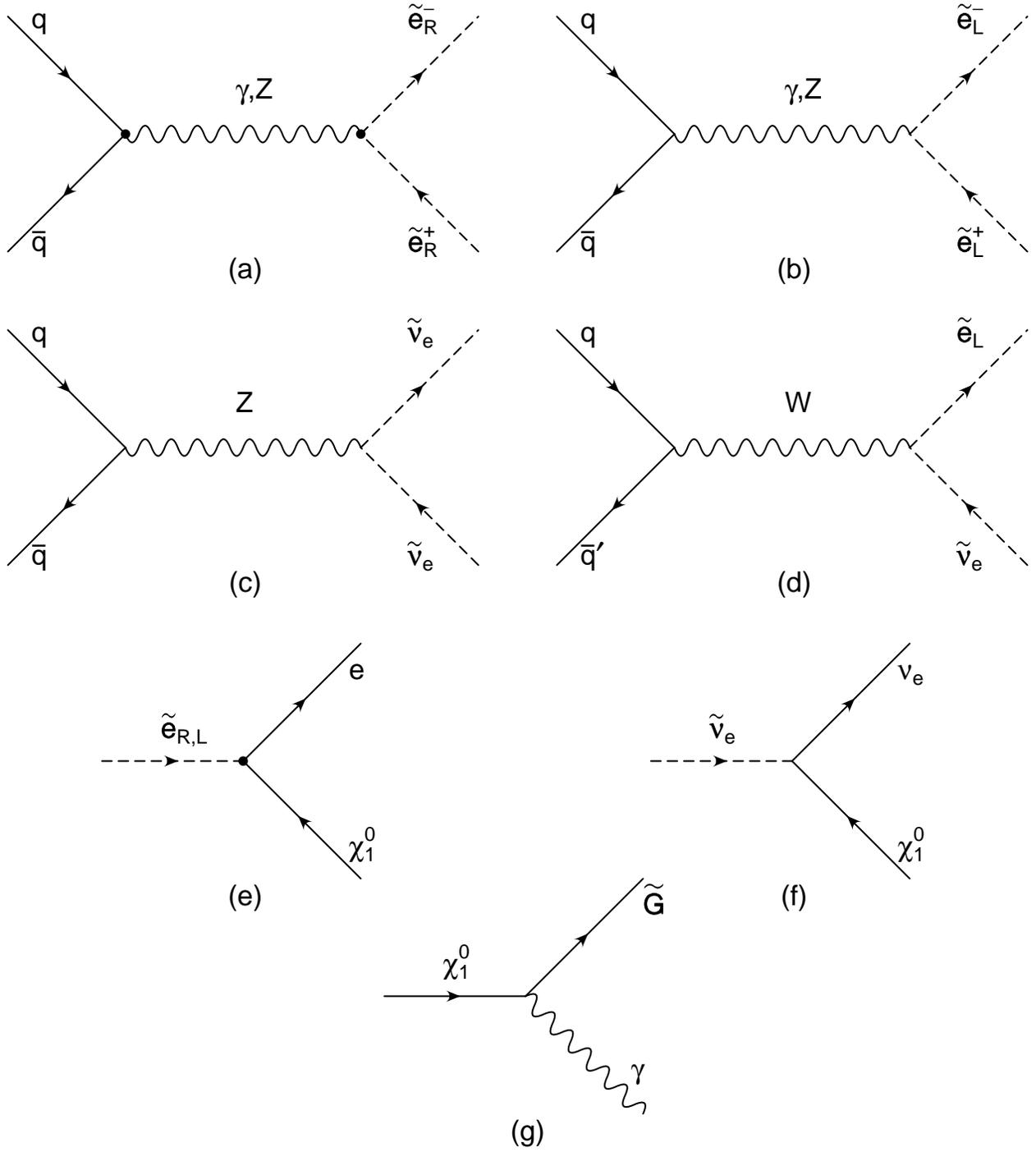}
\vspace{3cm}
\caption{Feynman diagrams for slepton production at the Tevatron, including
(a) $q\bar q\to\widetilde e^+_R\widetilde e^-_R$, 
(b) $q\bar q\to\widetilde e^+_L\widetilde e^-_L$, 
(c) $q\bar q\to\widetilde \nu_e\widetilde \nu_e$, 
(d) $q\bar q'\to\widetilde e^\pm_L\widetilde \nu_e$. Also shown are the dominant decay channels: (e) $\widetilde e_{R,L}\to e\chi^0_1$, (f) $\widetilde\nu_e\to\nu_e\chi^0_1$, and (g) $\chi^0_1\to\gamma\widetilde G$, with $\widetilde G$ representing the essentially massless gravitino. Analogous sets of diagrams exist for $\widetilde\mu,\widetilde\nu_\mu$ and $\widetilde\tau,\widetilde\nu_\tau$ production and decay.}
\label{fig:fd-sleptons}
\end{figure}

In this interpretation one assumes that the underlying process is 
$q\bar q\to\widetilde e^+_R\widetilde e^-_R$ or $q\bar q\to\widetilde e^+_L\widetilde e^-_L$, as shown in Figs.~\ref{fig:fd-sleptons}(a) and
\ref{fig:fd-sleptons}(b) respectively.\footnote{All Feynman diagrams in this
paper have been drawn using the software package {\tt FeynDiagram}, developed
by Bill Dimm (ftp://ftp.hepth.cornell.edu). We thank Jaewan Kim for procuring
and installing the package.} The selectrons subsequently decay via $\widetilde e^\pm_{R,L}\to e^\pm\chi^0_1$ [Fig.~\ref{fig:fd-sleptons}(e)]
with $100\%$ branching ratio,\footnote{In general the selectrons have several
possible decay channels: $\tilde e^\pm_R\to e^\pm\chi^0_1,e^\pm\chi^0_2$;
$\tilde e^\pm_L\to e^\pm\chi^0_1,e^\pm\chi^0_2,\nu_e\chi^\pm_1$. In the present model we find that $\tilde e^\pm_R\to e^\pm\chi^0_2$ is only accessible
for $m_{\tilde e_R}<70\,{\rm GeV}$ (and then suppressed by phase space), which
is outside of the range of interest. Also, $\tilde e_L$ may always
decay in all three ways (although $\tilde e^\pm_L\to e^\pm\chi^0_1$ has the
largest phase space), but with an electron in the final state almost always
(unless the chargino in $\tilde e_L\to\nu_e\chi^\pm_1$ decays hadronically).}
and the neutralinos further decay via
$\chi^0_1\to\gamma\widetilde G$  [Fig.~\ref{fig:fd-sleptons}(g)], also
with $100\%$ branching ratio. The final state
thus contains $e^+e^-\gamma\gamma\widetilde G\widetilde G$, with the (essentially) massless gravitinos carrying away the missing energy. 

\begin{figure}[p]
\vspace{6in}
\includegraphics{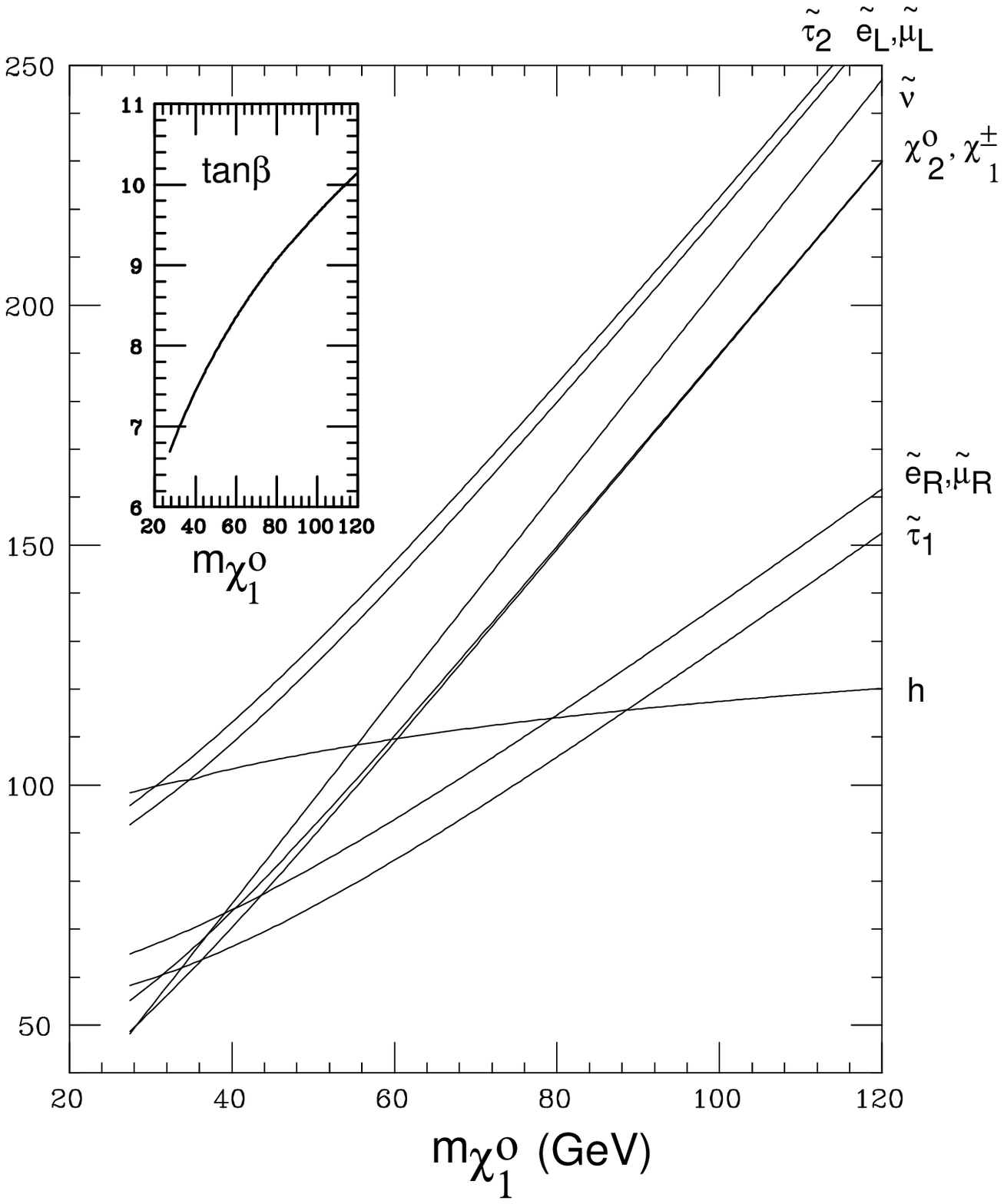}
\caption{The lighter members of the spectrum of our one-parameter model
versus the lightest neutralino mass. All masses in GeV. When multiple labels $a,b$ are attached to a particular line that may be split, the mass ordering
is $m_a\ge m_b$. The inset shows the variation of $\tan\beta$ with $m_{\chi^0_1}$.}
\label{fig:light}
\end{figure}

\begin{figure}[p]
\vspace{5in}
\includegraphics{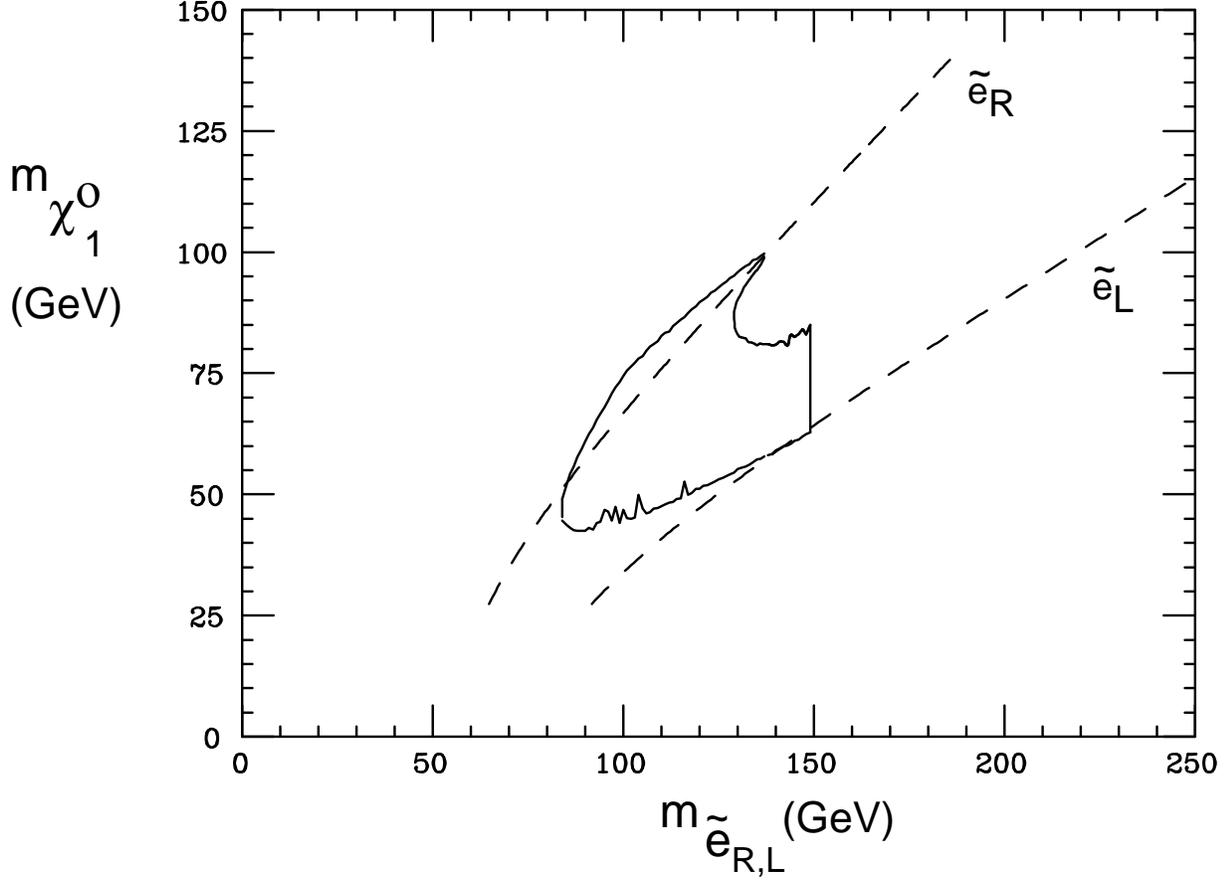}
\caption{Region consistent with the kinematics of the CDF $ee\gamma\gamma+E_T\hskip-13pt/\quad$ event when interpreted as selectron
($\widetilde e_{R,L}$) pair production: each selectron decays to electron
and lightest neutralino ($e\chi^0_1$) and each neutralino decays to photon and gravitino ($\gamma\widetilde G$). Region has been cutoff (vertical line)
where predicted rate becomes uninterestingly small. Kinematics disfavors $\widetilde e_L$ pair-production and is consistent with $m_{\tilde e_R}\approx(85-135)\,{\rm GeV}$ and $m_{\chi^0_1}\approx(50-100)\,{\rm GeV}$. Model mass relations then imply $m_{\chi^\pm_1}\approx(90-190)\,{\rm GeV}$.}
\label{fig:event1-sel}
\end{figure}

The kinematical data shown in Table~\ref{Table1} constrain the possible masses
of the selectron ($m_{\tilde e_{R,L}}$) and neutralino ($m_{\chi^0_1}$). To
obtain these constraints we perform a Monte Carlo simulation of the process
in which we vary at random the three-momenta of the unobserved gravitinos, subject to the constraints of transverse momentum conservation (up the $E_T\hskip-13pt/\quad$ `vector' given in Table~\ref{Table1}) and equality
of neutralino and selectron masses in each decay. We have considered the two
possible pairings of the photons and electrons ({\rm i.e.}, which $e$ and
$\gamma$ go with $\widetilde e^+$ and which go with $\widetilde e^-$)
but find that only one pairing satisfies all consistency constraints. We thus obtain a $2\times3-2-1-1=2$--dimensional solution space that parametrizes the two unknown masses. The resulting allowed region in $(m_{\tilde e_{R,L}},m_{\chi^0_1}$) space is found to be partially bounded, and with a sufficiently dense sample of Monte Carlo points one can determine its boundary, as shown in Fig.~\ref{fig:event1-sel}. The region in principle continues
onto larger sparticles masses, but we have cut it off (vertical line) at
$m_{\tilde e}=150\,{\rm GeV}$ because the predicted rates at the Tevatron become uninterestingly small beyond that point. Within the closed region
we see that underlying $\widetilde e_L$ pair-production is disfavored. On the other hand, underlying $\widetilde e_R$ pair-production is perfectly consistent with the kinematics of the event, with the further constraints
\begin{equation}
m_{\tilde e_R}\approx(85-135)\,{\rm GeV}\ ,\quad
m_{\chi^0_1}\approx(50-100)\,{\rm GeV}\ .
\label{eq:selconstraints}
\end{equation}
Model mass relations (see Fig.~\ref{fig:light}) then imply $m_{\chi^\pm_1}\approx(90-190)\,{\rm GeV}$.

\begin{table}[t]
\caption{The kinematical information of the observed CDF $ee\gamma\gamma+E_T\hskip-13pt/\quad$ event. All momenta and energies in GeV.
Also important are $E_T\hskip-13pt/\quad=52.81\,{\rm GeV}$ at $\phi=2.91\,{\rm rad}$.}
\label{Table1}
\begin{center}
\begin{tabular}{crrrr}
Variable&$e_1\quad$&$e_2\quad$&$\gamma_1\quad$&$\gamma_2\quad$\\ \hline
$p_x$&$58.75$&$-33.41$&$-12.98$&$31.53$\\
$p_y$&$18.44$&$11.13$&$-29.68$&$-17.48$\\
$p_z$&$-167.24$&$21.00$&$-22.69$&$-34.77$\\
$E$&$178.21$&$41.00$&$39.55$&$50.09$\\
$E_T$&$61.58$&$35.21$&$32.39$&$36.05$\\ \hline
\end{tabular}
\end{center}
\hrule
\end{table}

\begin{figure}[p]
\vspace{6in}
\includegraphics{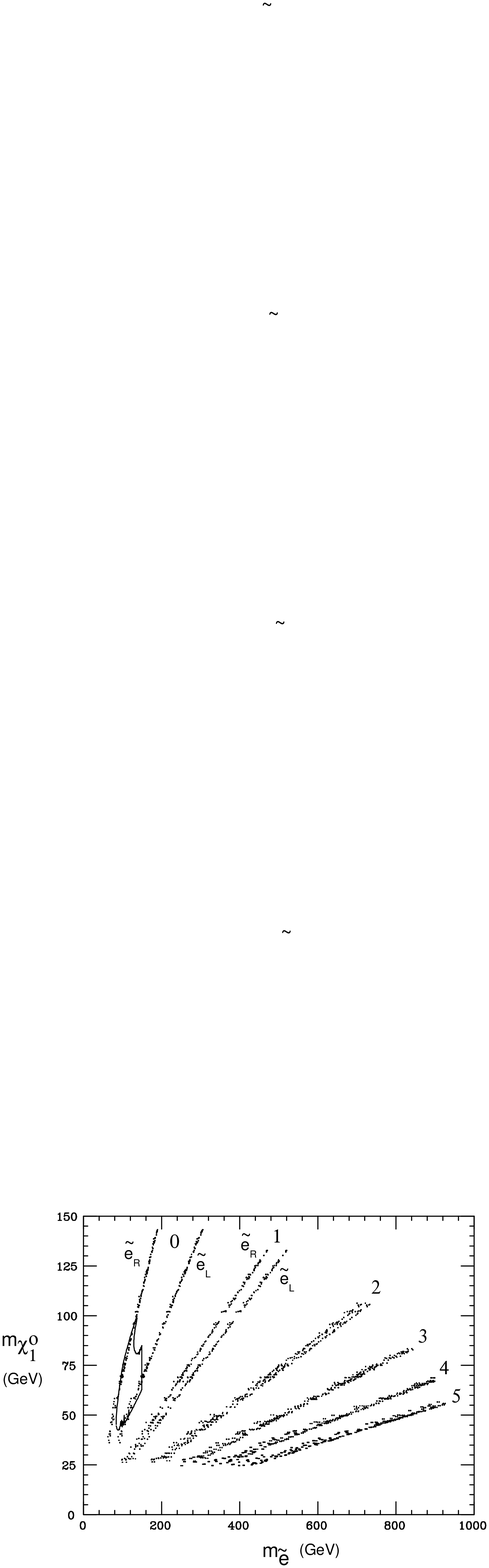}
\caption{Calculated distribution of selectron ($\tilde e$) and lightest
neutralino ($\chi^0_1$) masses in generic supergravity models for
fixed values of the ratio $\xi_0=m_0/m_{1/2}=0,1,2,3,4,5$; and varying values
of $\{m_{1/2}, \tan\beta,A_0\}$. The area within the closed region is consistent with the kinematics and dynamics of the CDF $ee\gamma\gamma$ event. The branches for $\tilde e_R$ and $\tilde e_L$ are only distinguishable for $\xi_0=0,1$. Our model predicts $\xi_0=0$.}
\label{fig:lspsel}
\end{figure}

Before proceeding with the analysis, it is instructive to determine what
other sets of standard supergravity parameters $\{m_0,m_{1/2},A_0,\tan\beta\}$
may also be consistent with the kinematics of the CDF event in the selectron
interpretation. We have generated 10,000 different random four-parameter sets of this kind, and in each case determined the low-energy spectrum, in particular the $\chi^0_1$ and $\tilde e_{R,L}$ masses. As we just showed,
the kinematics of the event singles out an allowed region in the
$(m_{\tilde e},m_{\chi^0_1})$ plane (see Fig.~\ref{fig:event1-sel}).
In Fig.~\ref{fig:lspsel} we show the distribution of models in this space, with the preferred region bounded by the solid line. For clarity, in the figure we restrict the choices of $\xi_0=m_0/m_{1/2}$ to the integer values shown ({\em i.e.}, $0\to5$), with the other three parameters allowed to vary at random. (The branches for $\tilde e_R$ and $\tilde e_L$ are only distinguishable for  $\xi_0=0,1$.) This figure illustrates the fraction of the generic supergravity parameter space that is consistent with the kinematics of the CDF event. Moreover, our model prediction of $\xi_0=0$ clearly falls within the allowed region for both $\tilde e_R$ and $\tilde e_L$, whereas $\xi_0\ge1$ is not allowed. Note that the actual model prediction for $\widetilde e_L$
(as shown in Fig.~\ref{fig:event1-sel}) has much less of an overlap with the
allowed region than the generic $m_0=0$ result in Fig.~\ref{fig:lspsel}. This
is due to the additional model constraints that correlate all four parameters,
leaving only one to vary freely.

\begin{figure}[p]
\vspace{6in}
\includegraphics{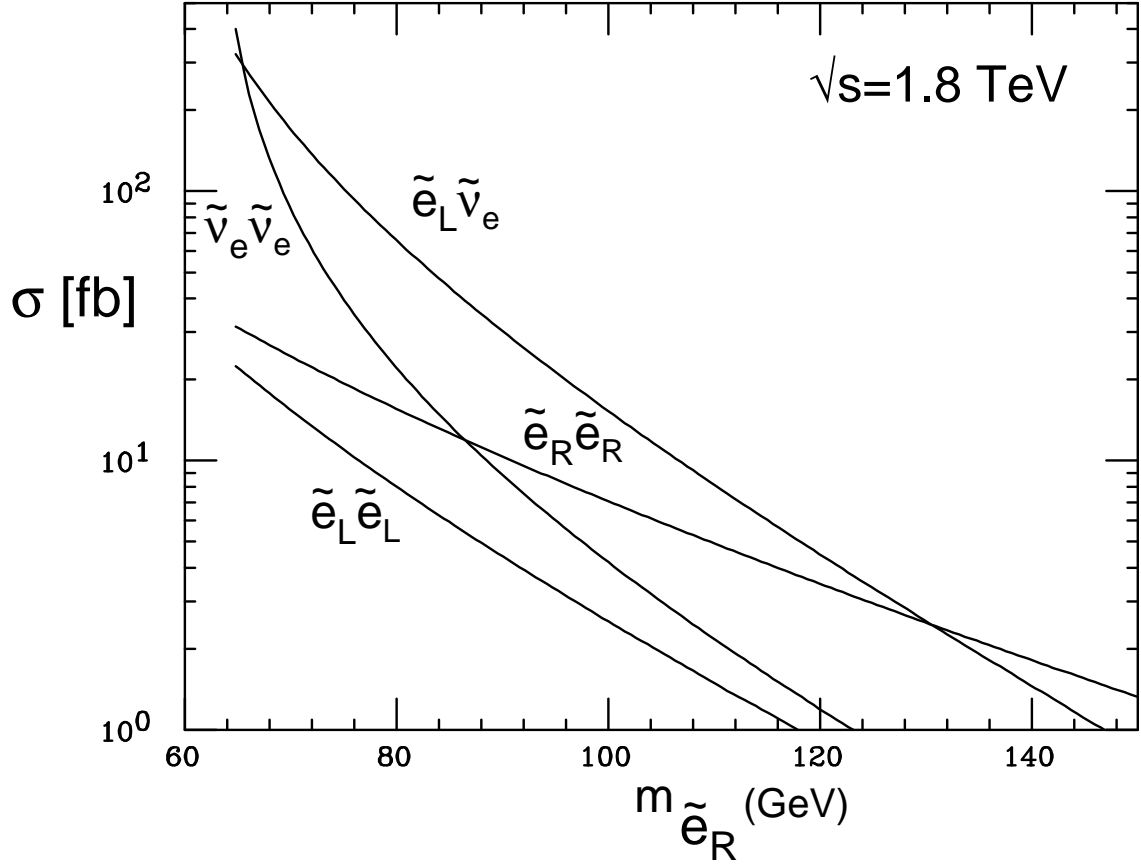}
\caption{Cross sections (in fb) for production of sleptons at the Tevatron
as a function of the selectron mass ($\widetilde e_R$). These final
states entail the following signatures: 
$e^+e^-\gamma\gamma+E_T\hskip-13pt/\quad
(\tilde e_{R,L}\tilde e_{R,L})$, 
$e^\pm\gamma\gamma+E_T\hskip-13pt/\quad(\tilde e_L\tilde\nu_e)$,
$\gamma\gamma+E_T\hskip-13pt/\quad(\tilde\nu_e\tilde\nu_e)$. The same curves
and analogous signatures hold for $\tilde\mu,\tilde\nu_\mu$ and $\tilde\tau,\tilde\nu_\tau$ production.}
\label{fig:signals-slep}
\end{figure}

We now turn to the dynamics of the selectron interpretation. We consider the
production of the four slepton channels: 
\begin{eqnarray}
&&q\bar q\to \gamma,Z\to\widetilde\ell^+_R\widetilde\ell^-_R,\widetilde \ell^+_L\widetilde\ell^-_L\to (\ell^+\chi^0_1)(\ell^-\chi^0_1) 
\to \ell^+\ell^-\gamma\gamma+E_T\hskip-13pt/\quad\
\label{eq:LLRR}\\
&&q\bar q'\to W^\pm\to \widetilde \ell^\pm_L\widetilde\nu_\ell\to
(\ell^\pm\chi^0_1)(\nu_\ell\chi^0_1)
\to\ell^\pm\gamma\gamma+E_T\hskip-13pt/\quad,
\label{eq:NuLL}\\
&&q\bar q\to Z\to\widetilde\nu_\ell\widetilde\nu_\ell\to (\nu_\ell\chi^0_1)(\nu_\ell\chi^0_1)
\to\gamma\gamma+E_T\hskip-13pt/\quad,
\label{eq:NuNu}
\end{eqnarray}
where $\ell=e,\mu,\tau$ [see Figs.~\ref{fig:fd-sleptons}(a,b,c,d)], and we
have indicated the experimental signature in each case ($E_T\hskip-13pt/\quad$
includes the two gravitinos from $\chi^0_1\to\gamma\widetilde G$ decay and
may also include neutrinos). We have calculated the corresponding hadronic cross sections using the parton-level differential cross sections given in Ref.~\cite{BCPT}, and integrated them over phase space and parton distribution functions. The resulting cross sections for $\ell=e$ are shown in Fig.~\ref{fig:signals-slep} as a function of the selectron mass. The same result is obtained for $\ell=\mu$ or $\ell=\tau$, {\em i.e.}, one would multiply the curves in  Fig.~\ref{fig:signals-slep} by a factor of 3 to obtain the cross section summed over the three flavors. We should note that because the slepton ($\widetilde\ell_R,\widetilde\ell_L,\widetilde\nu_\ell$) masses are correlated in the specific way shown in Fig.~\ref{fig:light}, the resulting relations between the different slepton rates differ from those naively obtained when the slepton masses of the same flavor are taken to be all equal. For instance, when $m_{\tilde e_L}=m_{\tilde e_R}$ one finds $\sigma(\tilde e_L\tilde e_L)\approx2.3\,\sigma(\tilde e_R\tilde e_R)$, whereas in our case
$\sigma(\tilde e_L\tilde e_L)<\sigma(\tilde e_R\tilde e_R)$ always.

\begin{figure}[p]
\vspace{6in}
\includegraphics{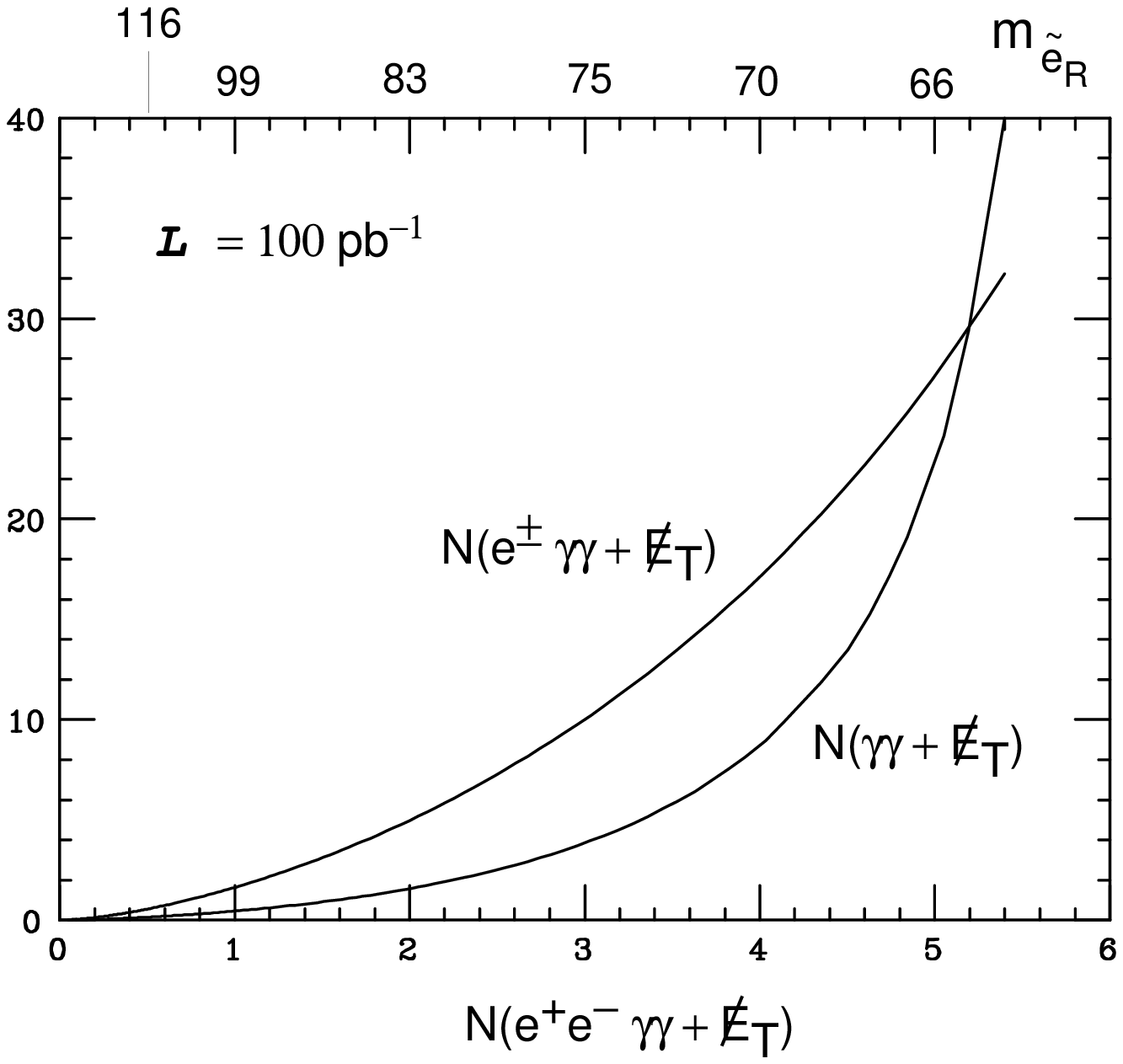}
\caption{Number of $e^\pm\gamma\gamma+E_T\hskip-13pt/\quad$
and $\gamma\gamma+E_T\hskip-13pt/\quad$ events versus the number of
$e^+e^-\gamma\gamma+E_T\hskip-13pt/\quad$ events expected from slepton production at the Tevatron in ${\cal L}=100\,{\rm pb}^{-1}$ of data. The
top axis shows the corresponding selectron masses ($\tilde e_R$) in GeV.}
\label{fig:1evs2e}
\end{figure}

As noted in Eqs.~(\ref{eq:LLRR},\ref{eq:NuLL},\ref{eq:NuNu}), the experimental
signatures for these four channels differ in the number of charged leptons:
2,1,0 respectively. In Fig.~\ref{fig:1evs2e} we show the correlation between
the number of single-lepton versus the number of di-lepton events that are
expected in ${\cal L}=100\,{\rm pb}^{-1}$ of data. At the one dilepton-event level, the expected number of single-lepton events (two) is still consistent with observation (zero). However, possible observation of more $ee\gamma\gamma+E_T\hskip-13pt/\quad$ events would need to be accompanied by many more $e\gamma\gamma+E_T\hskip-13pt/\quad$ events. On the top axis
of the plot we show the corresponding selectron mass, which shows that
one event is consistent with $m_{\tilde e_R}$ as high as 100 GeV. (However,
one should realize $N(ee\gamma\gamma+E_T\hskip-13pt/\quad)$ in Fig.~\ref{fig:1evs2e} includes both $\tilde e_R$ and $\tilde e_L$ sources, although the former dominates over the latter and the latter is kinematically
inconsistent with the observed event.) It has been advocated that $1\over2$ event should mark the minimum event rate consistent with the observations \cite{Dine,Kane}, in this case we find the upper bound
$m_{\tilde e_R}<115\,{\rm GeV}$. This limit may be weakened to
$m_{\tilde e_R}<150\,{\rm GeV}$ by summing over slepton flavors, although it is not clear what the meaning of this procedure is.

In Fig.~\ref{fig:1evs2e} we also show the number of diphoton (no-lepton) events, which is seen to be negligible at the one dilepton-event level, but should quickly overtake the number of dilepton events should more of these be observed. As a final comment on the viability of the selectron interpretation, we should remark that one must require a rather substantive chargino mass ($m_{\chi^\pm_1}>125\,{\rm GeV}$ \cite{Kane}) so that the additional sources
of $ee\gamma\gamma+E_T\hskip-13pt/\quad$ events from chargino/neutralino
production (see Sec.~\ref{sec:chargino}) remain below the few-event level. Through the mass relations in this model (see Fig.~\ref{fig:light}), this requirement entails $m_{\tilde e_R}>100\,{\rm GeV}$ and $m_{\chi^0_1}>68\,{\rm GeV}$, which are still consistent with the selectron interpretation.

\subsection{Chargino interpretation}
\label{sec:chargino}
\begin{figure}[p]
\vspace{6in}
\includegraphics{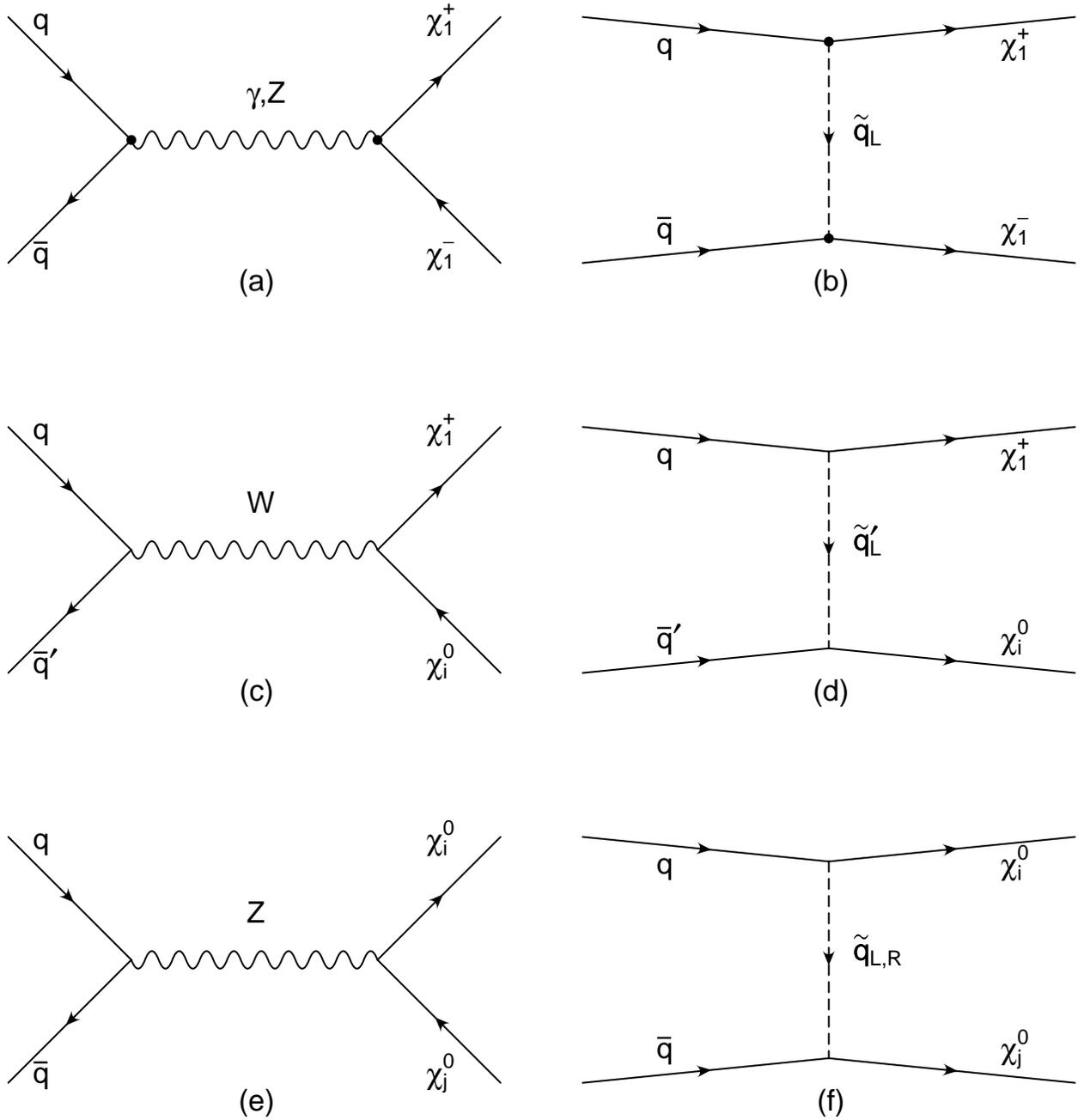}
\vspace{3cm}
\caption{Feynman diagrams for chargino/neutralino production at the Tevatron, including
(a,b) $q\bar q\to\chi^+_1\chi^-_1$, (c,d) $q\bar q'\to\chi^\pm_1\chi^0_i$, 
(e,f) $q\bar q\to\chi^0_i\chi^0_j$, where $i,j=1,2$. The squark-exchange
diagrams (b,d,f) are small. The $s$-channel neutralino production diagram (e)
is suppressed for gaugino-like neutralinos.}
\label{fig:fd-cn}
\end{figure}

\begin{figure}[p]
\vspace{6in}
\includegraphics{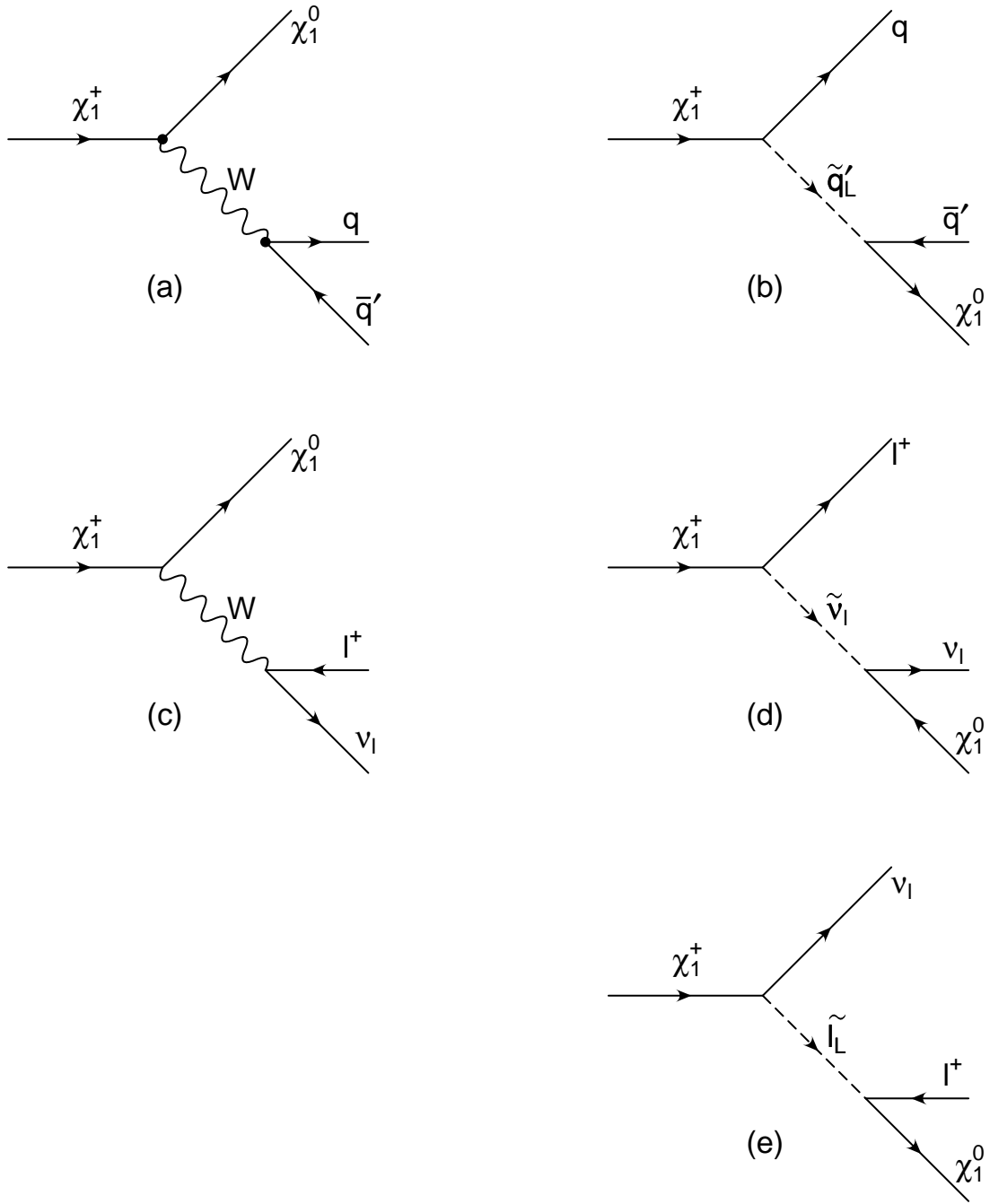}
\vspace{3cm}
\caption{Feynman diagrams for chargino decay: (a,b) 
$\chi^\pm_1\to q\bar q'\chi^0_1$, (c,d,e) $\chi^\pm_1\to\ell\nu_\ell^\pm\chi^0_1$ ($\ell=e,\mu,\tau$).}
\label{fig:fd-chdecay}
\end{figure}

One may envision an alternative interpretation of the $ee\gamma\gamma+E_T\hskip-13pt/\quad$ event wherein the underlying process is 
assumed to be $q\bar q\to\chi^+_1\chi^-_1$ as shown in Figs.~\ref{fig:fd-cn}(a,b). The charginos subsequently decay via
$\chi^\pm_1\to e^\pm\nu_e\chi^0_1$ [see Fig.~\ref{fig:fd-chdecay}(c,d,e)] 
with some calculable branching ratio, and the neutralinos further decay via
$\chi^0_1\to\gamma\widetilde G$  [Fig.~\ref{fig:fd-sleptons}(g)]
with $100\%$ branching ratio. The final state thus contains $e^+e^-\gamma\gamma\nu_e\bar\nu_e\widetilde G\widetilde G$, with the (essentially) massless gravitinos and the neutrinos carrying away the missing energy. 

\begin{figure}[p]
\vspace{5in}
\includegraphics{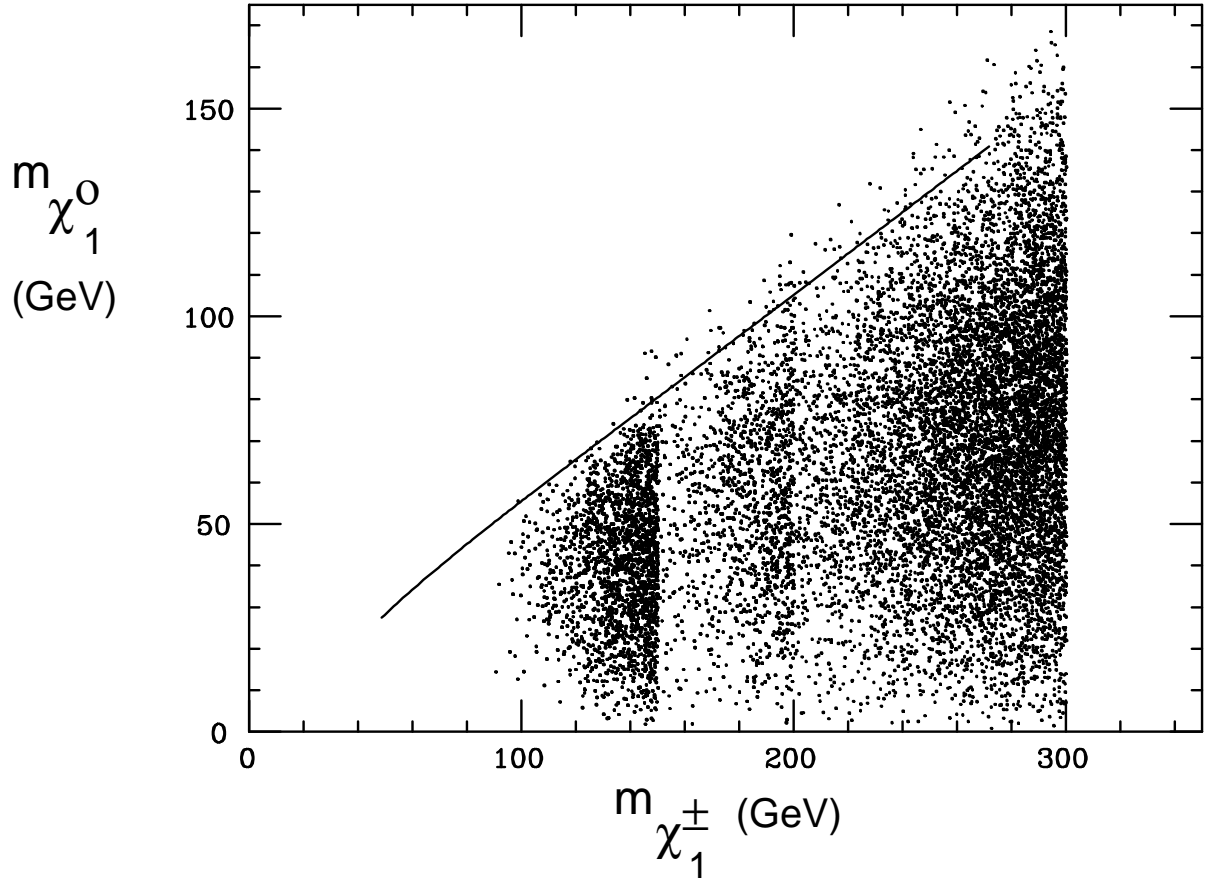}
\caption{Region consistent with the kinematics of the CDF $ee\gamma\gamma+E_T\hskip-13pt/\quad$ event when interpreted as chargino
($\chi^\pm_1$) pair production: each chargino decays to $e\nu_e\chi^0_1$ and each neutralino decays to photon and gravitino ($\gamma\widetilde G$). Kinematics requires $m_{\chi^\pm_1}>95\,{\rm GeV}$. The solid line represents the model prediction, which is consistent with the kinematics of the event for
$m_{\chi^\pm_1}>100\,{\rm GeV}$ and $m_{\chi^0_1}>55\,{\rm GeV}$.}
\label{fig:event1-ch}
\end{figure}

In analogy with the selectron interpretation, the kinematical data shown in Table~\ref{Table1} constrain the possible masses of the chargino ($m_{\chi^\pm_1}$) and neutralino ($m_{\chi^0_1}$). (These constraints hold
irrespective of the dynamics of the process, {\em i.e.}, the underlying cross
section and branching ratios.) However, in the chargino interpretation we
have four missing momenta, and the constraints from tranverse momentum
conservation and equality of neutralino and chargino masses in each decay
allow a vast $4\times3-2-1-1=8$--dimensional solution space. We find that both
possible pairings of the photons and electrons ({\em i.e.}, which $e$ and
$\gamma$ go with $\chi^+_1$ and which go with $\chi^-_1$) satisfy all consistency constraints. The resulting allowed region in $(m_{\chi^\pm_1},m_{\chi^0_1}$) space is difficult to delineate, but as far
as we can tell it is partially bounded requiring $m_{\chi^\pm_1}>95\,{\rm GeV}$, as shown in Fig.~\ref{fig:event1-ch}. The figure also shows that 
the model prediction (solid line) does intercept the kinematically allowed
region, starting at $m_{\chi^\pm_1}\approx100\,{\rm GeV}$ and $m_{\chi^0_1}\approx55\,{\rm GeV}$. The dynamics of the event further restricts the allowed interval, as we discuss below.

We would like to digress briefly to comment on other sets of standard supergravity parameters $\{m_0,m_{1/2},A_0,\tan\beta\}$ that may also be consistent with the kinematics of the event in the chargino interpretation. Following the same procedure as in the selectron interpretation, we find the well-known result in supergravity models that $m_{\chi^\pm_1}\approx 2m_{\chi^0_1}$. Therefore, basically {\em any} standard supergravity model that allows $m_{\chi^\pm_1}>100\,{\rm GeV}$ will be consistent with the kinematical constraints of the event in the chargino interpretation. Of course the dynamics of the event (light gravitino, suitable cross section and branching ratios) will be a lot more discriminating. For instance, any value of $m_0$ would appear acceptable from the kinematics alone. However, $m_0$ affects the branching ratios and may not enhance the desired leptonic decays of the chargino. Also, as discussed in Sec.~\ref{sec:motivation}, a non-negligible value of $m_0$ is not attainable in supergravity models with a light gravitino.

\begin{figure}[p]
\vspace{6in}
\includegraphics{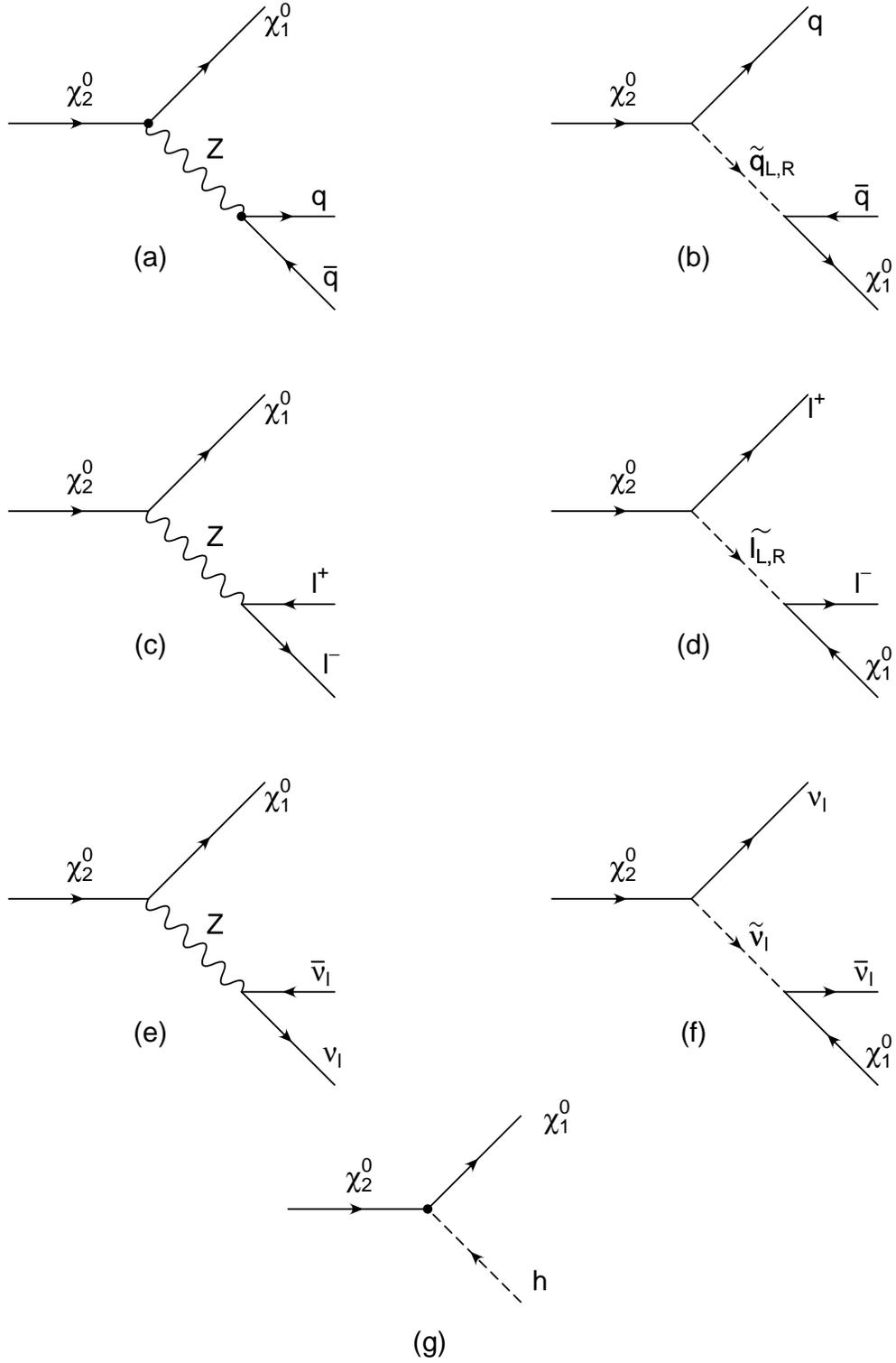}
\vspace{5cm}
\caption{Feynman diagrams for neutralino decay: (a,b) 
$\chi^0_2\to q\bar q\chi^0_1$, (c,d) $\chi^0_2\to\ell^+\ell^-\chi^0_1$,
(e,f) $\chi^0_2\to\nu_\ell\bar\nu_\ell\chi^0_1$ ($\ell=e,\mu,\tau$),
(g) $\chi^0_2\to h\chi^0_1$.}
\label{fig:fd-ndecay}
\end{figure}

We now turn to the dynamics of the chargino interpretation. We consider the
production of the following chargino/neutralino channels: 
\begin{eqnarray}
&&q\bar q\to \chi^+_1\chi^-_1\to
\left\{\begin{array}{l}
(\ell^+\nu_\ell\chi^0_1)(\ell^-\bar\nu_\ell\chi^0_1)
\to\ell^+\ell^-\gamma\gamma+E_T\hskip-13pt/\\
(\ell^\pm\nu_\ell\chi^0_1)(q\bar q'\chi^0_1)
\to \ell^\pm\,jj\,\gamma\gamma+E_T\hskip-13pt/\\
(q\bar q'\chi^0_1)(q'\bar q\chi^0_1)
\to jjjj\,\gamma\gamma+E_T\hskip-13pt/\quad
\end{array}\right.
\label{eq:C1C1}\\
&&q\bar q'\to\chi^\pm_1\chi^0_2\to
\left\{\begin{array}{l}
(\ell^\pm\nu_\ell\chi^0_1)(\ell^{'+}\ell^{'-}\chi^0_1)
\to \ell^\pm\ell^{'+}\ell^{'-}\,\gamma\gamma+E_T\hskip-13pt/\\
(\ell^\pm\nu_\ell\chi^0_1)(\nu_\ell\bar\nu_\ell\chi^0_1)
\to \ell^\pm\,\gamma\gamma+E_T\hskip-13pt/\\
(q\bar q'\chi^0_1)(\ell^+\ell^-\chi^0_1)
\to \ell^+\ell^-\,jj\,\gamma\gamma+E_T\hskip-13pt/\\
(\ell^\pm\nu_\ell\chi^0_1)(q\bar q\chi^0_1)
\to \ell^\pm\,jj\,\gamma\gamma+E_T\hskip-13pt/\\
(q\bar q'\chi^0_1)(\nu_\ell\bar\nu_\ell\chi^0_1)
\to jj\,\gamma\gamma+E_T\hskip-13pt/\\
(q\bar q'\chi^0_1)(q\bar q\chi^0_1)
\to jjjj\,\gamma\gamma+E_T\hskip-13pt/\\
\end{array}\right.
\label{eq:C1N2}\\
&&q\bar q'\to\chi^\pm_1\chi^0_1\to
\left\{\begin{array}{l}
\chi^0_1(\ell^\pm\nu_\ell\chi^0_1)\to \ell^\pm\,\gamma\gamma+E_T\hskip-13pt/\\
\chi^0_1(q\bar q'\chi^0_1)\to jj\,\gamma\gamma+E_T\hskip-13pt/\\
\end{array}\right.
\label{eq:C1N1}
\end{eqnarray}
where $\ell,\ell'=e,\mu,\tau$, and we have indicated the experimental signature in each case. We have not shown the $\chi^0_2\to h\chi^0_1$ decay channel [see Fig.~\ref{fig:fd-ndecay}(g)] as in our model it becomes kinematically accessible only for $m_{\chi^0_2}>m_h+m_{\chi^0_1}$ or $m_{\chi^0_2}>2 m_h\approx250\,{\rm GeV}$. Also, the additional production channels
$q\bar q\to\chi^0_i\chi^0_j$ ($i,j=1,2$) shown in Fig.~\ref{fig:fd-cn}(e,f),
proceed at much smaller rates: the $Z\chi^0_i\chi^0_j$ coupling is suppressed for our case of mostly gaugino-like neutralinos, and the $t$-channel diagrams
are suppressed by heavy squark masses.

\begin{figure}[p]
\vspace{5in}
\includegraphics{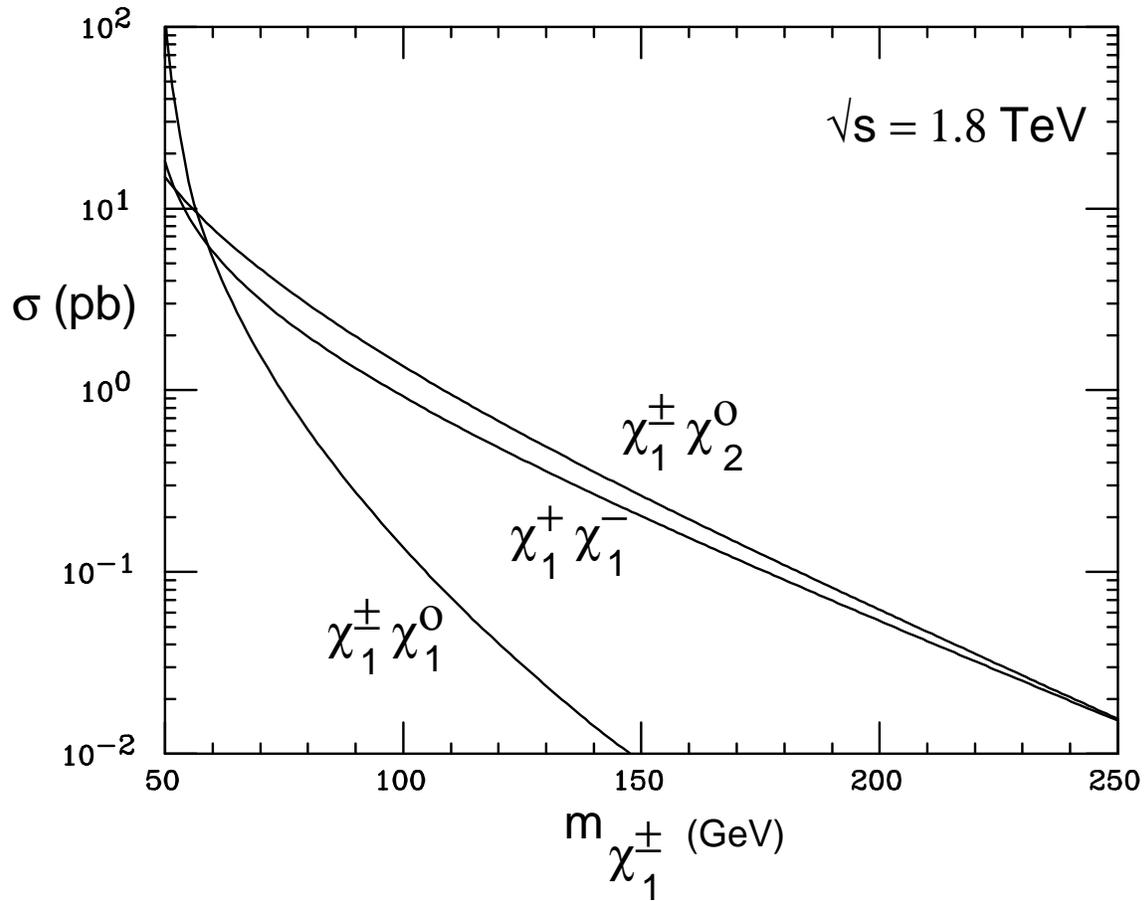}
\caption{The cross sections for the dominant chargino/neutralino production
processes at the Tevatron versus the chargino mass. }
\label{fig:fermi}
\end{figure}

We have calculated the hadronic cross sections for the processes in Eqs.~(\ref{eq:C1C1},\ref{eq:C1N2},\ref{eq:C1N1}), and these are shown in Fig.~\ref{fig:fermi} as a function of the chargino mass. We note that for not too small values of the chargino mass, the $\chi^\pm_1\chi^0_2$ and
$\chi^+_1\chi^-_1$ channels dominate. However, $\chi^\pm_1\chi^0_1$ is not negligible and may give the dominant contribution to some signatures, as we discuss below.  

\begin{figure}[p]
\vspace{6in}
\includegraphics{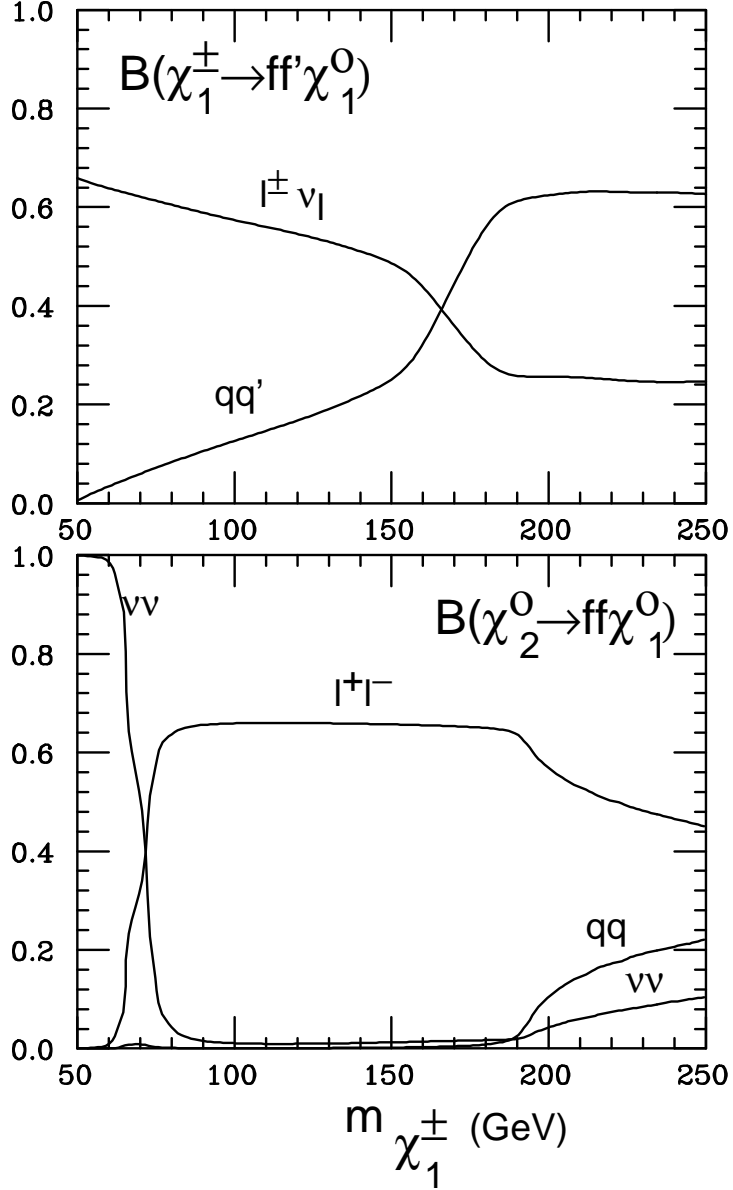}
\caption{The chargino ($\chi^\pm_1$) and neutralino ($\chi^0_2$) branching ratios as a function of the chargino mass (note that $m_{\chi^\pm_1}\approx m_{\chi^0_2}$). The $\ell^\pm\nu_\ell$ and $\ell^+\ell^-$ branching ratios include $\ell=e+\mu$.}
\label{fig:br-ratios}
\end{figure}

The chargino and neutralino branching ratios are complicated functions of all model parameters, as the Feynman diagrams in Figs.~\ref{fig:fd-chdecay} and \ref{fig:fd-ndecay} indicate. The numerical results as a function of the chargino mass are shown in Fig.~\ref{fig:br-ratios}, and may be understood
as follows. In the case of the chargino decays, for low values of the chargino
mass the $\ell^\pm\nu_\ell$ channel dominates, as the sneutrino-mediated
decay [Fig.~\ref{fig:fd-chdecay}(d)] is enhanced by $m_{\tilde\nu}\approx m_{\chi^\pm_1}$ [see Fig.~\ref{fig:light}]. As the chargino mass increases the $W$-mediated decays become more important as the sneutrino becomes heavier than the chargino. Eventually the $W$-mediated decays fully dominate, and the expected branching ratios into leptons ($3/9$) and into quarks ($6/9$) are attained. (Note that in Fig.~\ref{fig:br-ratios} we have summed over only two lepton families.) 

In the case of neutralino decays, the pattern may also be understood in terms of the spectrum shown in Fig.~\ref{fig:light}. For low neutralino masses, the
sneutrino-mediated decay [Fig.~\ref{fig:fd-ndecay}(f)] is on-shell and $\chi^0_2\to\nu_\ell\bar\nu_\ell\chi^0_1$ dominates. Soon thereafter the
sneutrino goes off-shell, but the right-handed sleptons go on-shell and
$\chi^0_2\to\ell^+\ell^-\chi^0_1$ dominates.\footnote{This leptonic decay
dominance is not present in the chargino decay because the chargino does not
couple to the right-handed sleptons, which are the ones that mediate this
dominant amplitude.} This dominance continues until the decay $\chi^0_2\to Z\chi^0_1$ becomes accessible for $m_{\chi^0_2}\approx2M_Z$ (as Fig.~\ref{fig:br-ratios} shows). The influence of the $Z$-mediated channels
continues until $m_{\chi^0_2}\approx 2m_h$, where the $\chi^0_2\to h\chi^0_1$ channel becomes accessible and completely dominant (such high values of $m_{\chi^0_2}$ are not shown in Fig.~\ref{fig:br-ratios}).

\begin{figure}[p]
\vspace{6in}
\includegraphics{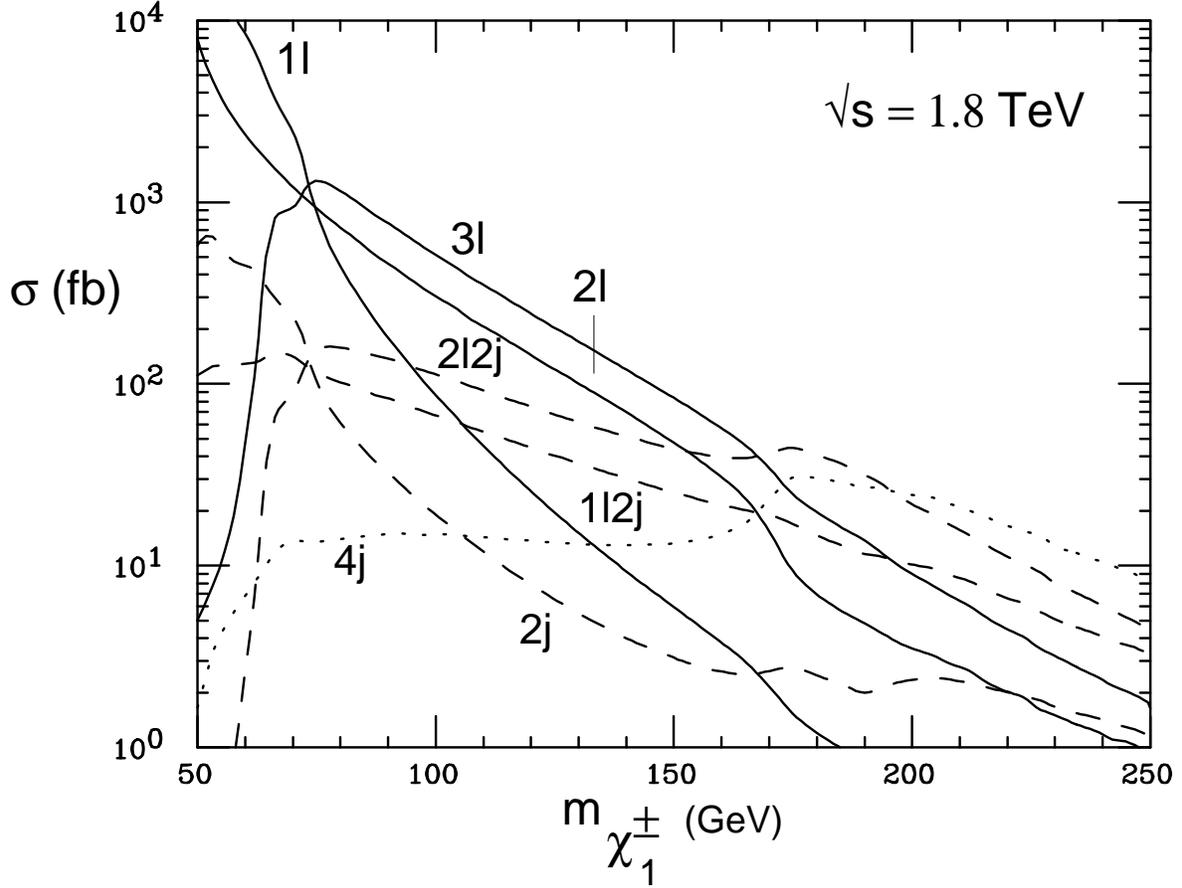}
\caption{The rates for the various ${\rm n}\ell\,{\rm m}j$ signals (with n charged leptons and m jets) obtained from chargino/neutralino
production at the Tevatron versus the chargino mass. All indicated signals
are accompanied by $\gamma\gamma+E_T\hskip-13pt/\quad$. The leptonic signals
have been summed over $\ell=e+\mu$. One may estimate the number of expected events in $100\,{\rm pb}^{-1}$ of data to be $1/100$ of the indicated rates by assuming an experimental detection efficiency of $10\%$. Note that in the
region of interest [$m_{\chi^\pm_1}\approx(100-150)\,{\rm GeV}$] events without jets (solid lines) dominate over events with 2 jets (dashed lines) or events with 4 jets (dotted line).}
\label{fig:signals-cn}
\end{figure}

The actual signals are obtained by multiplying the cross sections times the
branching ratios as indicated in Eqs.~(\ref{eq:C1C1},\ref{eq:C1N2},\ref{eq:C1N1}); these are shown in
Fig.~\ref{fig:signals-cn}. This busy plot shows that the expected number of events without jets (solid lines) dominates over the number of events with 2 jets (dashed lines) or 4 jets (dotted line). The precise number of
expected events depends on the experimental detection efficiencies for each
of these signatures, and thus requires a detailed simulation including detector
effects. One may get an idea of the number of expected events in ${\cal L}=100\,{\rm pb}^{-1}$ by simply assuming an across-the-board $10\%$ efficiency. In this approximation the numbers of expected events would be the
numbers in Fig.~\ref{fig:signals-cn} multiplied by $1/100$. These numbers of
expected events should not exceed a few, as so far only one event of this
kind has been observed at CDF. Consulting Fig.~\ref{fig:signals-cn}, it would
appear reasonable to require $m_{\chi^\pm_1}>100\,{\rm GeV}$ (which implies
$m_{\tilde e_R}>85\,{\rm GeV}$ and $m_{\chi^0_1}>55\,{\rm GeV}$). As discussed above (see Fig.~\ref{fig:event1-ch}), this lower bound is also required by the kinematics of the event. A reasonable upper bound in order to produce enough events would
appear to be $m_{\chi^\pm_1}\approx150\,{\rm GeV}$. Figure\ref{fig:signals-cn} then shows that in this regime the dominant
signals are $\ell^\pm\ell^{'+}\ell^{'-}+\gamma\gamma+E_T\hskip-13pt/\quad$
(including $e^\pm e^+e^-,e^+ e^-\mu^\pm,
e^\pm\mu^+\mu^-,\mu^\pm\mu^+\mu^-$),
$\ell^+\ell^- +\gamma\gamma+E_T\hskip-13pt/\quad$ (including $e^+e^-,\mu^+\mu^-$), and 
$\ell^+\ell^-jj +\gamma\gamma+E_T\hskip-13pt/\quad$ (including $e^+e^-jj,\mu^+\mu^-jj$). Note that the single-flavor $3\ell$ and $2\ell$ signals ({\em e.g.}, $e^\pm e^+e^-$ and $e^+e^-$) occur at nearly the same rate, as does the $\ell^+\ell^-jj$ signal given the uncertainty on the acceptances for these channels. (Decays involving $\ell=\tau$ are of course
also expected to occur.)

Let us also remark on the importance of the $\chi^\pm_1\chi^0_1$ channel in
Eq.~(\ref{eq:C1N1}), which contributes to the $1\ell$ and $2j$ signals in
Fig.~\ref{fig:signals-cn}. Both these signals receive an additional contribution from $\chi^\pm_1\chi^0_2$, but only when the neutralino decays invisibly ($\chi^0_2\to\nu\bar\nu\chi^0_1$), which happens only for very light chargino masses [see Fig.~\ref{fig:br-ratios}]. For moderate chargino masses these signals receive contributions only from the otherwise unimportant
$\chi^\pm_1\chi^0_1$ channel.

\section{LEP constraints and prospects}
\label{sec:LEP}
We now turn to the experimental consequences of our model that may be observable at LEP~2. In either selectron or chargino interpretation, we
have found that $m_{\tilde e_R}>80\,{\rm GeV}$ and more likely $m_{\tilde e_R}>100\,{\rm GeV}$ rendering the sleptons ($\tilde\ell_{R,L},\tilde\nu_\ell$) kinematically inaccessible at any LEP~2 energy presently being considered
({\em i.e.}, $\sqrt{s}=161,175,190\,{\rm GeV}$). The same goes for the charginos, where $m_{\chi^\pm_1}>100\,{\rm GeV}$ is required. The only possibly observable channels are then $\chi^0_1\chi^0_1$ and perhaps also $\chi^0_1\chi^0_2$, with the former having a maximum neutralino mass reach of $\sqrt{s}/2$ and the latter of approximately $\sqrt{s}/3$ (using $m_{\chi^0_2}\approx 2m_{\chi^0_1}$).

\subsection{Neutralino production}
\begin{figure}[p]
\vspace{3in}
\includegraphics{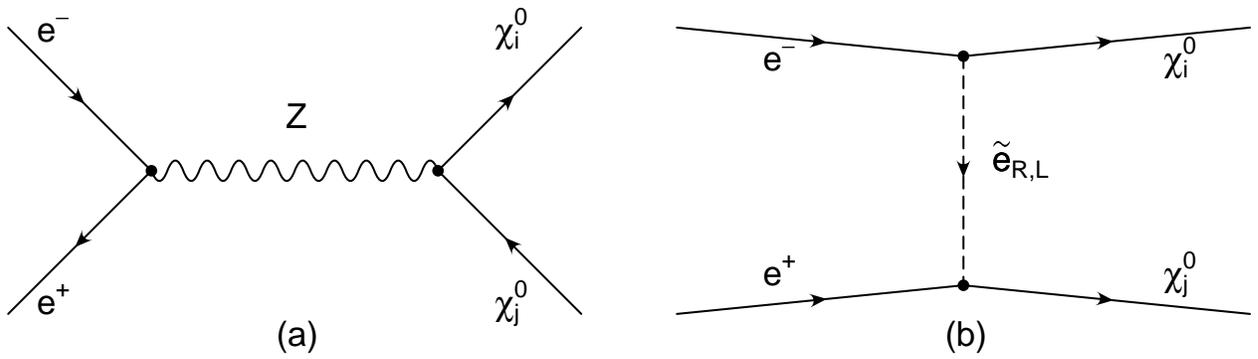}
\caption{Feynman diagrams for neutralino production at LEP. The right-handed
selectron ($\tilde e_R$) exchange $t$-channel diagram dominates for gaugino-like neutralinos.}
\label{fig:fd-NiNj}
\end{figure}

\begin{figure}[p]
\vspace{6in}
\includegraphics{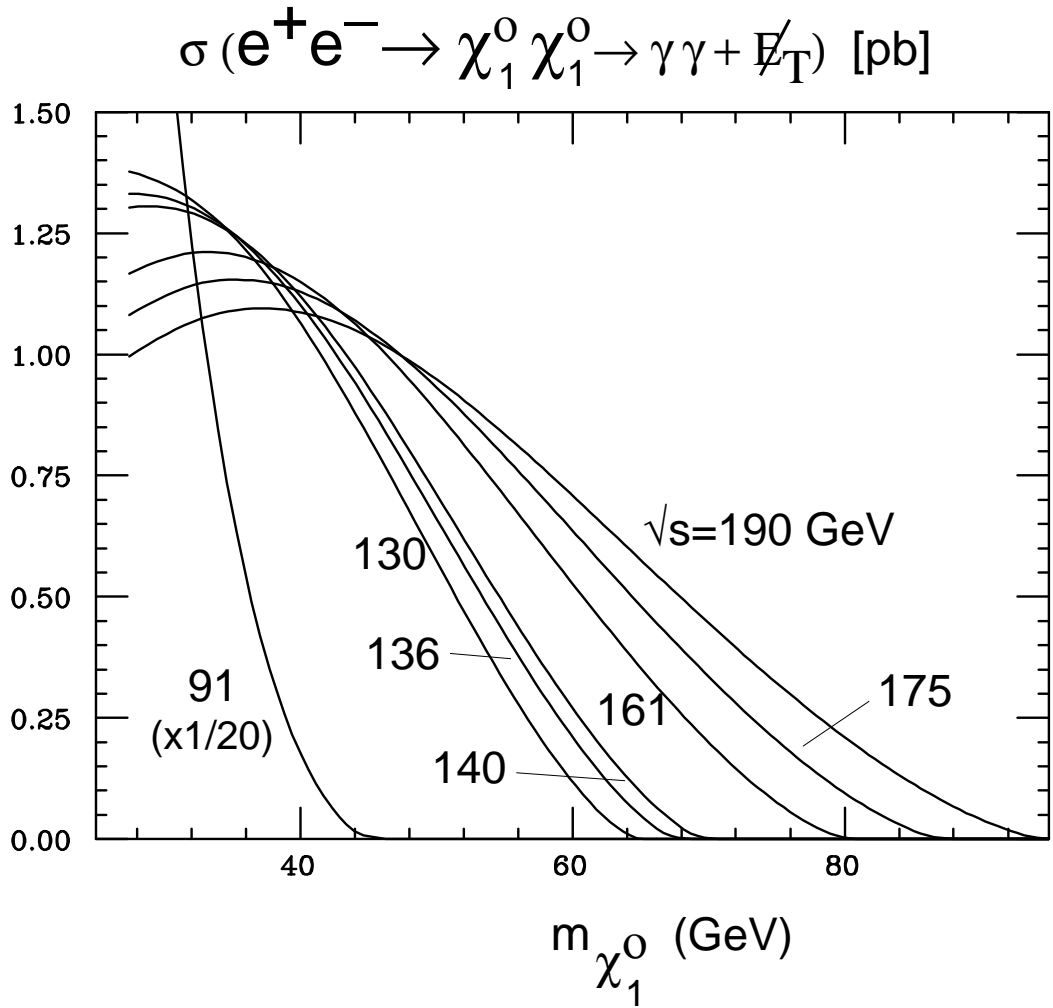}
\caption{The calculated cross section for lightest neutralino ($\chi^0_1$) production at LEP versus the neutralino mass for several center-of-mass energies.}
\label{fig:N1N1}
\end{figure}

The neutralino production processes proceed via $s$-channel $Z$ exchange and $t$-channel selectron exchange, as shown in Fig.~\ref{fig:fd-NiNj}.
The calculated cross section for the dominant $\chi^0_1\chi^0_1$ mode is shown in Fig.~\ref{fig:N1N1} as a function of the neutralino mass for several center-of-mass energies. This figure reveals several interesting features. First of all, our neutralinos are not exactly
pure binos: their relatively small higgsino admixture decreases with increasing neutralino mass [see discussion after Eq.~(\ref{eq:width})]. This admixture is responsible for the significant cross section at the $Z$ peak via the $s$-channel exchange diagram in Fig.~\ref{fig:fd-NiNj}(a). The higgsino admixture is also responsible for a non-negligible destructive interference between the $s$- and $t$-channel diagrams (analogous to that in chargino pair production) for light neutralino masses: dropping the $s$-channel diagram we obtain a cross section twice as large. For heavier neutralino masses this effect diminishes quickly, but a new effect turns on, namely the $\beta^3$ threshold dependence expected for P-wave production of neutralinos \cite{EH}.
We should also remark that the $\tilde e_R$ exchange diagram dominates
over the $\tilde e_L$ exchange diagram because $m_{\tilde e_R}<m_{\tilde e_L}$,
and more importantly because the coupling of selectrons to binos is proportional to the selectron hypercharge: $|Y_{\tilde e_R}/Y_{\tilde e_L}|=2$.

\begin{figure}[p]
\vspace{6in}
\includegraphics{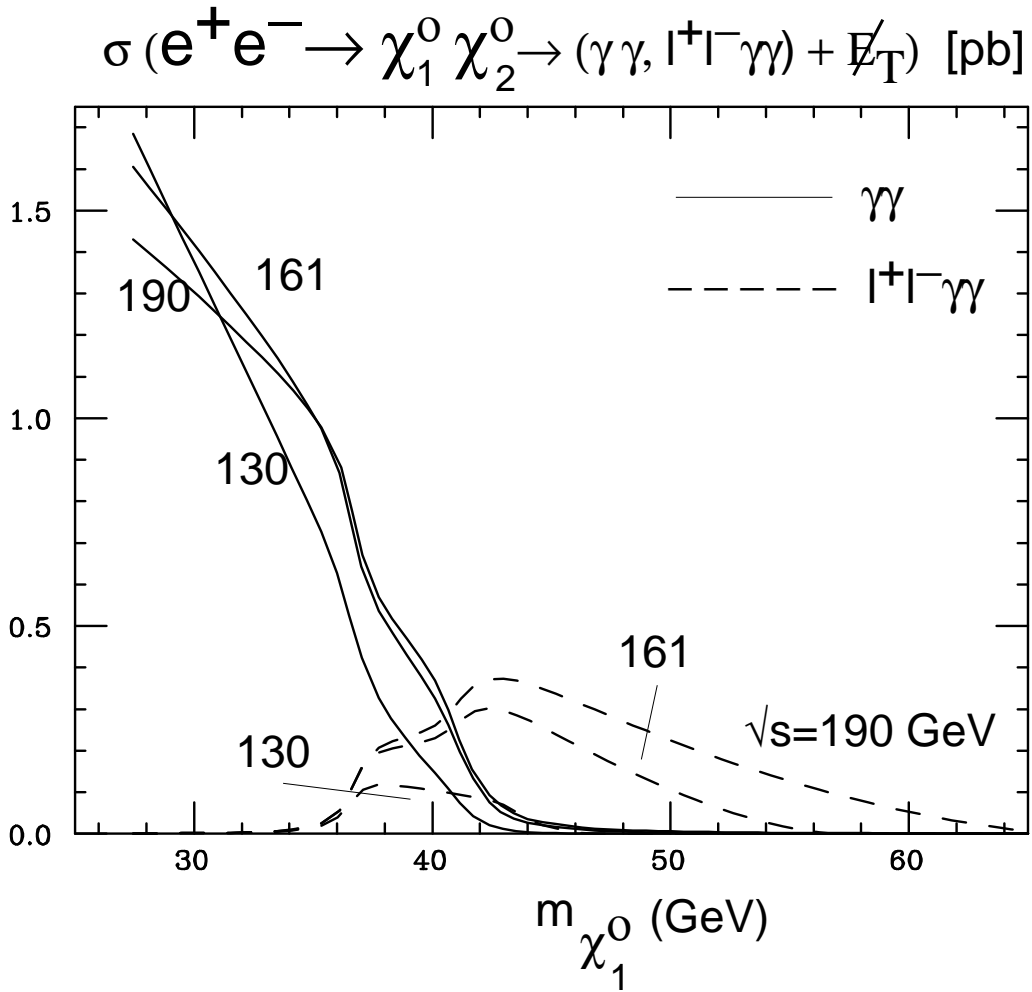}
\caption{The calculated cross section for $e^+e^-\to\chi^0_1\chi^0_2$ production at LEP times the corresponding $\chi^0_2$ branching ratio versus the neutralino mass for selected center-of-mass energies. }
\label{fig:N1N2}
\end{figure}

The only other channel of interest is $\chi^0_1\chi^0_2$, which leads to the
following signals
\begin{equation}
e^+e^-\to\chi^0_1\chi^0_2\to\left\{\begin{array}{l}
\chi^0_1(\nu_\ell\bar\nu_\ell\chi^0_1)\to\gamma\gamma+E_T\hskip-13pt/\\
\chi^0_1(\ell^+\ell^-\chi^0_1)\to\ell^+\ell^-\gamma\gamma+E_T\hskip-13pt/\\
\chi^0_1(q\bar q\chi^0_1)\to jj\,\gamma\gamma+E_T\hskip-13pt/
\end{array}\right.
\label{eq:N1N2}
\end{equation}
From the neutralino branching ratios in Fig.~\ref{fig:br-ratios} we can see that in the region of interest the dijet signal is negligible, while the other two signals do not occur simultaneously. The resulting cross sections times branching ratios are shown in Fig.~\ref{fig:N1N2}.

\begin{figure}[p]
\vspace{6in}
\includegraphics{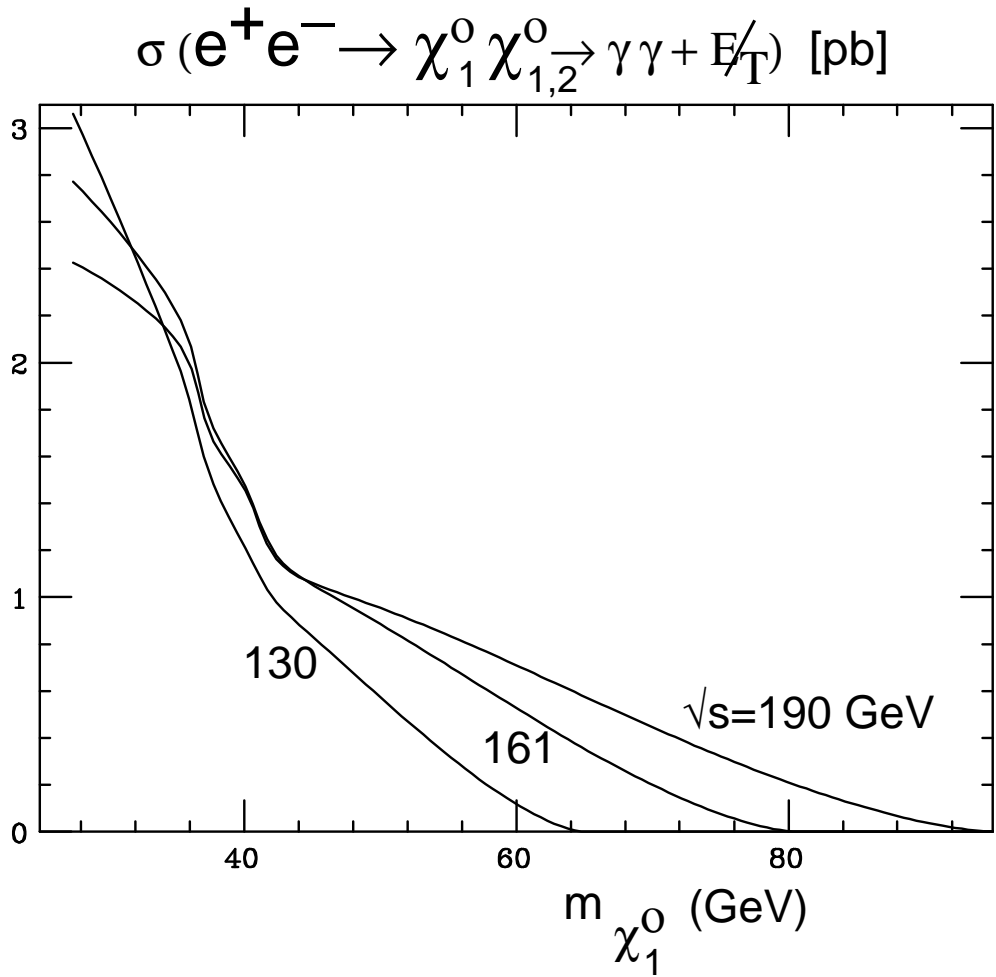}
\caption{The total $\gamma\gamma+E_T\hskip-13pt/\quad$ signal at selected
LEP energies versus the neutralino mass. This signal receives contributions
from the $\chi^0_1\chi^0_1$ and $\chi^0_1\chi^0_2$ channels. The kinks in the
curves reflect the $\chi^0_1\chi^0_2$ contribution going to zero.}
\label{fig:diphotons}
\end{figure}

Experimental constraints on beyond-the-standard-model contributions to acoplanar photon pairs at LEP~1.5 have been released by OPAL \cite{OPAL} and DELPHI \cite{DELPHI}, and they amount to upper bounds of 2.0~pb and 1.5~pb respectively. It is not clear whether these limits apply to only the $\chi^0_1\chi^0_1$ mode, or to the total $\gamma\gamma+E_T\hskip-13pt/\quad$ signal (that includes also contributions from $\chi^0_1\chi^0_2$). Moreover, these experimental limits are subject to angular cuts ($|\cos\theta|<0.7$) that we have not imposed on our signal. To be conservative we have applied these upper bounds to our total $\gamma\gamma+E_T\hskip-13pt/\quad$ signal. 
The total diphoton cross section is shown in Fig.~\ref{fig:diphotons}, from which we see that at LEP~1.5 it exceeds 1.5~pb for $m_{\chi^0_1}<37\,{\rm GeV}$ (corresponding to $m_{\chi^\pm_1}<65\,{\rm GeV}$), which may then be excluded (subject to the caveats just mentioned).
Note the kinks in the curves due to the $\chi^0_1\chi^0_2$ contribution going to zero (see Fig.~\ref{fig:N1N2}). DELPHI has also quoted an upper bound on the beyond-the-standard-model diphoton cross section at $\sqrt{s}=161\,{\rm GeV}$ of 3.1~pb. This limit does not exclude any region of parameter space; it would have to be strengthened by a factor of 2 before it starts to exclude points in parameter space not excluded by LEP~1.5. This may be the case with OPAL, which
has obtained an upper bound of 1.6~pb at $\sqrt{s}=161\,{\rm GeV}$ \cite{Wilson}.

\subsection{OPAL events}
\label{sec:OPAL}
Even though runs at LEP~130-136 and at LEP~161 have not revealed any clear
evidence for the diphoton signal that we have discussed above, it is possible
that the desired signal may occur only rarely and then surrounded by considerable background events. The OPAL Collaboration \cite{OPAL} in its
LEP~1.5 analysis identified six events with acoplanar photon pairs that warrant
closer scrutiny to see if the kinematical information of the events may
favor a signal rather than a background explanation. The particulars of these
events are listed in Table~\ref{Table2}. As OPAL noted, the first five events,
with a missing invariant mass $M_{\rm miss}\approx M_Z$, appear quite likely
to be from $e^+e^-\to Z\to\nu\bar\nu$, with an on-shell $Z$ boson obtained by the two photons realizing the ``radiative return to the Z" scenario. A Monte Carlo simulation of the distribution of $M_{\rm miss}$ in this kind of background events shows that it indeed peaks near $M_Z$, with a sharp low-end cutoff at $M_{\rm miss}\approx80\,{\rm GeV}$ \cite{Kane}. The diphoton signal on the other hand has an essentially flat distribution in $M_{\rm miss}$. We are then led to consider the last event in Table~\ref{Table2} as a possible signal event. 

\begin{figure}[p]
\vspace{6in}
\includegraphics{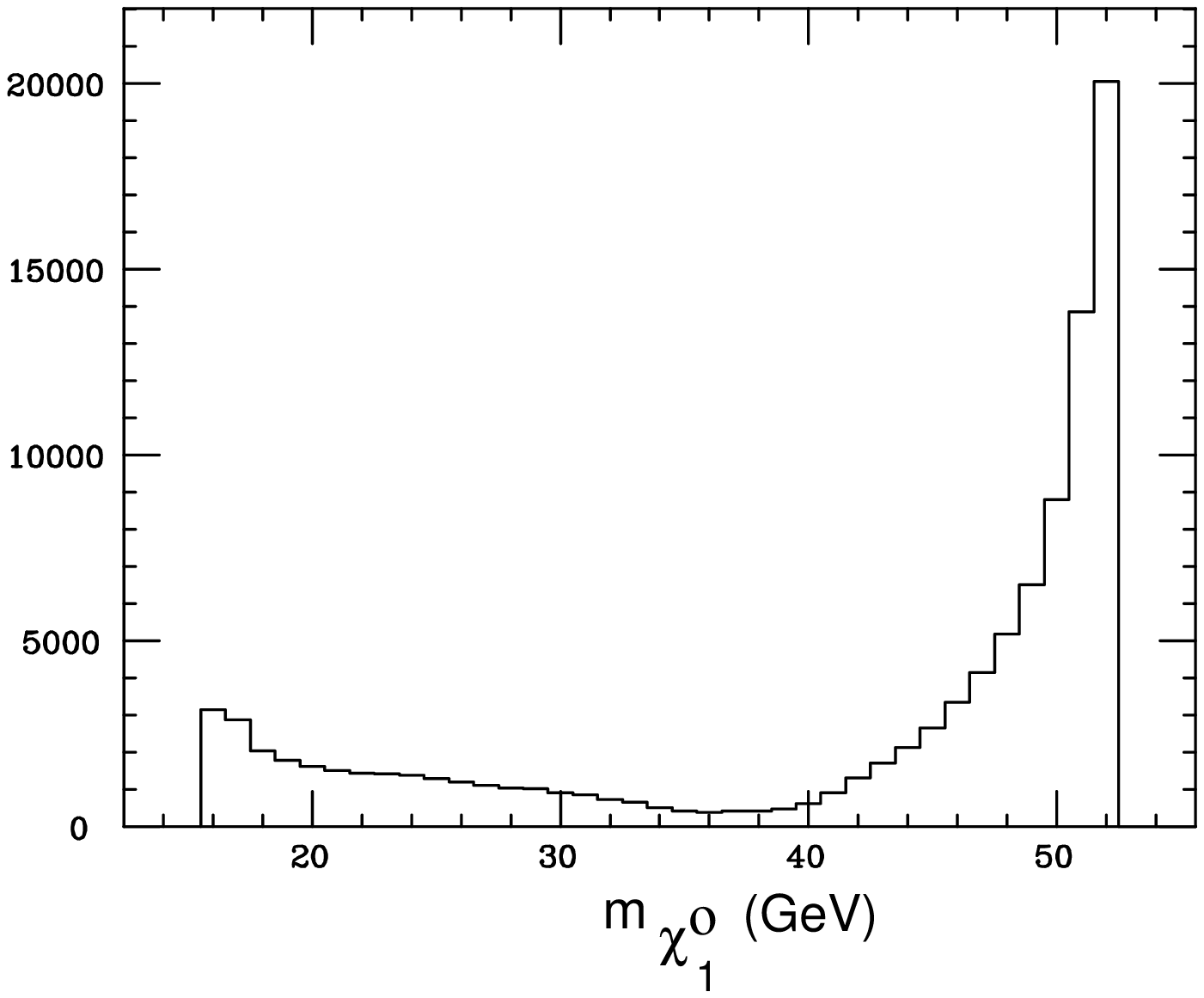}
\caption{Distribution of neutralino masses obtained by assuming that one of the
acoplanar photon pairs observed by OPAL originates from $e^+e^-\to\chi^0_1\chi^0_1\to\gamma\gamma+E_T\hskip-13pt/\quad$.}
\label{fig:bins}
\end{figure}

The kinematical information on the chosen event can be used to obtain constraints on the neutralino mass as follows. The event is assumed to be
$e^+e^-\to\chi^0_1\chi^0_1$ with subsequent decay $\chi^0_1\to\gamma\widetilde G$. In analogy with the exercises performed in Secs.~\ref{sec:selectron} and \ref{sec:chargino}, we vary the two three-momenta of the unobserved gravitinos and impose energy and momentum conservation plus
equality of the decaying neutralino masses. This gives us a $2\times3-1-3-1=1$
dimensional solution space, with the one degree of freedom parametrizing the
possible values of $m_{\chi^0_1}$. This exercise yields the distribution shown in Fig.~\ref{fig:bins}. One can see that a finite range of neutralino
masses $m_{\chi^0_1}\approx(17-53)\,{\rm GeV}$ is consistent with the kinematics of the event, and perhaps more useful, the distribution is strongly
peaked near its upper end. One could then assume that the
event may have come from neutralino production with $m_{\chi^0_1}\approx 52\,{\rm GeV}$ (corresponding to $m_{\chi^\pm_1}\approx100\,{\rm GeV}$).
The cross section at this mass at $\sqrt{s}=130\,{\rm GeV}$ is 0.45~pb (see Fig.~\ref{fig:N1N1}), and with an integrated luminosity of
${\cal L}=2.64\,{\rm pb}^{-1}$ one would expect 1.2 events. Taking into account a reasonable detection efficiency ($35\%$), we see that this event is quite
consistent with the kinematics and dynamics of the signal.

This interpretation of the OPAL event is also consistent with the lower range of the chargino interpretation of the CDF event. However, it appears inconsistent with the corresponding selectron interpretation, once the constraints from non-excessive $e^+e^-\gamma\gamma+E_T\hskip-13pt/\quad$ events from chargino production are taken into account (see end of Sec.~\ref{sec:selectron}).

Repeating the above exercise for the remaining events in Table~\ref{Table2}
yields similarly shaped distributions which are peaked at lower neutralino masses, ranging between 18 GeV (fourth event) and 34 GeV (third event). These
events are quite consistent with the missing mass ($M_{\rm miss}$) expected from the background and, in any event, would correspond to neutralino masses which have been already excluded by LEP~1.5 ($m_{\chi^0_1}>37\,{\rm GeV}$). 

\begin{table}[t]
\caption{The kinematical information of the six acoplanar photon pair events observed by OPAL at LEP~1.5. Also shown is the missing invariant mass $M_{\rm miss}$. All momenta, masses, and energies in GeV, angles in radians.}
\label{Table2}
\begin{center}
\begin{tabular}{ccrrrccc}
$\sqrt{s}$&$E_1$&$E_2$&$\cos\theta_1$&$\cos\theta_2$ &$\phi_1$&$\phi_2$
&$M_{\rm miss}$\\ \hline
$130.26$&$31.9$&$2.9$&$-0.312$&$-0.928$&$1.818$&$4.411$&$90.0\pm1.9$\\
$130.26$&$29.4$&$5.9$&$0.484$&$0.165$&$2.147$&$4.574$&$91.2\pm1.7$\\
$136.23$&$35.2$&$4.8$&$0.465$&$0.413$&$1.413$&$2.484$&$88.3\pm2.2$\\
$136.23$&$35.2$&$2.2$&$-0.230$&$-0.026$&$5.845$&$4.147$&$92.4\pm2.0$\\     
$136.23$&$36.1$&$2.4$&$-0.154$&$-0.505$&$4.104$&$4.799$&$90.0\pm2.2$\\
$130.26$&$28.5$&$18.4$&$0.473$&$-0.926$&$4.064$&$1.172$&$81.3\pm1.7$\\
\hline
\end{tabular}
\end{center}
\hrule
\end{table}

\subsection{Future acoplanar photon pairs}
In anticipation of future acoplanar photon pair events that may have 
$M_{\rm miss}<80\,{\rm GeV}$, we have determined analytically the end
points of the distribution of neutralino masses (as shown {\em e.g.}, in Fig.~\ref{fig:bins}), given the three-momenta of the observed photons. The neutralino mass is given by
\begin{equation}
m_{\chi^0_1}=2\sqrt{E_1\left({\sqrt{s}\over2}-E_1\right)}\,
\sin{\psi_{11'}\over2}\ ,
\label{eq:N1mass}
\end{equation}
where $E_1$ is the energy of the most energetic photon and $E_{1'}={\sqrt{s}\over2}-E_1$ is the energy of the accompanying gravitino. The angle $\psi_{11'}$ between these two particles is the one free parameter, which is however restricted by momentum conservation. This angle takes its minimum and maximum values when the momenta of the photon ($\vec p_1$) and the gravitino ($\vec p_{1'}$) are coplanar with the sum of the two photon momenta ($\vec p_1+\vec p_2$). This can be visualized by considering the plane formed by the two gravitino momenta ($\vec p_{1'},\vec p_{2'}$) and $\vec p_1+\vec p_2$. On this plane all vectors and their directions are completely determined by momentum conservation. However, the plane defined by $\vec p_1,\vec p_2$ may rotate around the $\vec p_1+\vec p_2$ axis. When the ($\vec p_1,\vec p_2$) plane is coplanar with the original  ($\vec p_{1'},\vec p_{2'},\vec p_1+\vec p_2$) plane, $\psi_{11'}$ attains its extremal values: 
\begin{equation}
\psi_{11'}^{\rm min}=\pi-\psi_{1'2'}-\psi_{1,1+2}\,,\qquad
\psi_{11'}^{\rm max}=\pi-\psi_{1'2'}+\psi_{1,1+2}\ ,
\label{eq:angles}
\end{equation}
where $\psi_{1'2'}$ is the angle between the gravitinos and $\psi_{1,1+2}$ is the angle between the most energetic photon and the sum of the photon momenta.
These angles can be determined from
\begin{equation}
\cos\psi_{1'2'}={p^2_{\rm miss}+E^2_{1'}-E^2_{2'}\over 2p_{\rm miss}E_{1'}}\,,
\quad
\cos\psi_{1,1+2}={p^2_{\rm miss}+E^2_{1}-E^2_{2}\over 2p_{\rm miss}E_{1}}\,,
\label{eq:psi's}
\end{equation}
with the missing momentum $p_{\rm miss}=|\vec p_1+\vec p_2|$, and the
gravitino energies $E_{1',2'}={\sqrt{s}\over2}-E_{1,2}$. For the sixth event
in Table~\ref{Table2} we find $\psi_{1'2'}=110.3^\circ$ and $\psi_{1,1+2}=39.5^\circ$, giving $30.2^\circ<\psi_{11'}<109.2^\circ$ and thus
$0.26<\sin{1\over2}\psi_{11'}<0.81$ (compared to the naive range $0<\sin{1\over2}\psi_{11'}<1$ obtained by neglecting momentum conservation). 

Should candidate acoplanar photon pair events be identified at LEP~2, the
above analysis would determine the range of consistent neutralino masses,
with their most likely value very near $m^{\rm max}_{\chi^0_1}$ ({\em i.e},
when $\psi_{11'}\approx\psi^{\rm max}_{11'}$).

\section{Conclusions}
\label{sec:conclusions}
Motivated by the puzzling CDF $ee\gamma\gamma+E_T\hskip-13pt/\quad$ event, we
have studied two different interpretations within the context of our recently
proposed one-parameter no-scale supergravity model with a light gravitino.
We considered both a selectron pair-production interpretation and a chargino pair-production interpretation. The former is consistent with the kinematics and dynamics of the event only for $\tilde e_R$, with some further constraints. In the context of this model we do not expect that more such events should turn up, unless many more similar events with a single lepton are also identified. The chargino interpretation can explain the CDF event with more ease and further predicts additional signals that may also be detectable at the same level (including events with three leptons and two leptons plus two jets). Possible observation of any of these accompanying signals will support the model. We also find that LEP~2 is in an ideal position to confirm or disprove the model via the detection of an excess of acoplanar photon pair events. In fact, we have noticed that one such event with characteristics more of a signal than of a background may have already been detected by OPAL at LEP~1.5

To conclude let us remark that in the case of light enough gravitinos, direct gravitino production at colliders ({\em e.g.}, $e^+e^-\to\chi^0_1\widetilde G
\to\gamma+E_T\hskip-13pt/\quad$) should provide an additional signal that would have a much greater kinematical reach \cite{Dicus,LNZ}.

\section*{Acknowledgments}
J. L. would like to thank Geary Eppley for useful discussions, Teruki Kamon for 
providing the information in Table~\ref{Table1}, Graham Wilson for providing the information in Table~\ref{Table2}, and Alan Litke for pointing out an error in the neutralino cross section at LEP. The work of J.~L. has been supported in part by the Associated Western Universities Faculty Fellowship program and in part by DOE grant DE-FG05-93-ER-40717. The work of D.V.N. has been supported in part by DOE grant DE-FG05-91-ER-40633.

\newpage
\clearpage

\clearpage

\clearpage

\clearpage

\clearpage

\clearpage

\clearpage

\clearpage

\clearpage

\clearpage

\clearpage

\clearpage

\clearpage

\clearpage

\clearpage

\clearpage

\clearpage

\end{document}